\documentstyle[epsf]{lamuphys}
\begin{document}

\title{Frustrated Systems: 
Ground State Properties via Combinatorial Optimization
\footnote{Lecture given on the E\"otv\"os summer school in Physics:\\
{\bf Advances in Computer Simulations},
Budapest, July 16--20, 1996.}
}

\author{Heiko Rieger}
\institute{HLRZ c/o Forschungszentrum J\"ulich, 52425 J\"ulich, Germany}

\titlerunning{Frustrated Systems: Ground State Properties}
\maketitle

\begin{abstract}
An introduction to the application of combinatorial optimization
methods to ground state calculations of frustrated, disordered systems
is given. We discuss the interface problem in the random bond Ising
ferromagnet, the random field Ising model, the diluted antiferromagnet
in an external field, the spin glass problem, the solid-on-solid
model with a disordered substrate and other convex cost flow problems
occurring in superconducting flux line lattices and traffic flow
networks. On the algorithmic side we present a concise introduction
to a number of elementary algorithms in combinatorial optimization, in
particular network flows: the shortest path algorithm, the
maximum-flow algorithms and minimum-cost-flow algorithms.
We present a short glance at the minimum weighted matching and
branch-and-cut algorithms.
\end{abstract}
\vskip3cm

\begin{center}
To be published in:\\
{\bf Lecture Notes in Physics}\\
(Springer-Verlag, Heidelberg-New$\,$York, 1997)
\end{center}

\vfill
\eject

\newcommand{\bc}{\begin{center}}
\newcommand{\ec}{\end{center}}
\newcommand{\be}{\begin{equation}}
\newcommand{\ee}{\end{equation}}
\newcommand{\beqn}{\begin{eqnarray}}
\newcommand{\eeqn}{\end{eqnarray}}
\newcommand{\ba}{\begin{array}}
\newcommand{\ea}{\end{array}}
\newcommand{\s}{{s}}
\renewcommand{\t}{{t}}
\renewcommand{\ss}{\sigma}

\newcommand{\spacea}{\rule{0.5cm}{0mm}}
\newcommand{\spaceb}{\rule{1.0cm}{0mm}}
\newcommand{\spacec}{\rule{1.5cm}{0mm}}
\newcommand{\spaced}{\rule{2.0cm}{0mm}}
\newcommand{\spacee}{\rule{2.5cm}{0mm}}

\noindent
{\underline{\bf Contents}}

\bc
\begin{enumerate}
\item What are frustrated systems?
\item What is special for computer simulations of disordered systems?
\item What can you learn from {\it ground state} calculations?
\item Ground state interface in a random medium.
\item The random field Ising model.
\item The diluted antiferromagnet in an external field.
\item The spin glass problem.
\item The SOS (solid-on-solid) model on a disordered substrate.
\item Vortex glasses and traffic flows.
\smallskip
\item[]\underline{Appendix:} Concepts in network flows and basic algorithms.\hfill\\
\begin{itemize}
\item[A] Maximum flow / minimum cut problem.
\begin{itemize}
\item[A.1] Basic definitions.
\item[A.2] Residual network and generic augmenting path algorithm.
\item[A.3] Cuts, labeling algorithm and max-flow-min-cut theorem.
\item[A.4] Generic Preflow-push algorithm.
\end{itemize}
\item[B] Shortest path problem.
\begin{itemize}
\item[B.1] Dijkstra's algorithm.
\item[B.2] Label correcting algorithm.
\end{itemize}
\item[C] Minimum cost flow problem.
\begin{itemize}
\item[C.1] Definition.
\item[C.2] Negative cycle canceling algorithm.
\item[C.3] Reduced cost optimality.
\item[C.4] Successive shortest path algorithm.
\item[C.5] Convex cost flows.
\end{itemize}
\end{itemize}
\end{enumerate}
\ec

\vfill
\eject

\section{What are frustrated systems?}

Frustrated systems are simply systems in which the individual
entities that build up the model (like spins, bosons, fermions,
monomers, etc.) feel some sort of ``frustration'' in the literal
sense. This means that on their search for a minimal energy
configuration at lower and lower temperatures they are not able to
satisfy all interactions with one another or with impurities
simultaneously.

As an example we consider a model for a directed polymer in a 
disordered environment
\be
H=\underbrace{\sum_i(x_i-x_{i+1})^2}_{\displaystyle\bf (A)}
-\underbrace{\sum_i V(x_i)}_{\displaystyle\bf (B)}\;,
\ee
where $x_i$ is the displacement of the $i$-th monomer and $V(x)$ is a
random potential. The first term (A), the elastic
energy, tries to make the polymer {\it straight} for $T\to0$, the
second term (B) tries to bring the monomers in favorite positions, for
which the polymer has to {\it bend}. The monomers cannot satisfy 
both of these demands simultaneously.

Another, more famous example are magnetic spins (for simplicity Ising
spins) with ferromagnetic {\it and} antiferromagnetic interactions.
Consider the Hamiltonian for 4 spins (e.g.\ an elementary plaquette of
a square lattice)
\be
H=-\ss_1\ss_2-\ss_2\ss_3-\ss_3\ss_4+\ss_4\ss_1
\label{fourspins}
\ee 
and try to find a configuration of the Ising spins $\ss_i=\pm1$
that minimizes this simple energy function. Naively one starts with
some value for the first spin, let's say $\ss_1=+1$, the first term
would then imply $\ss_2=+1$, the second $\ss_3=+1$ and the third
$\ss_4=+1$. But what about the last term --- here $\ss_1=\ss_4+1$ is
{\it not} the most favorable configuration. Thus it is impossible to
satisfy all local interactions at once, this is why
Toulouse\cite{toulouse} introduced the concept of frustration for
these plaquette occurring naturally in spin glass models. After some
thought one finds that many (i.e.\ 8) different spin configurations
for (\ref{fourspins}) have a minimal energy, but all of them break one
bond. This is the notorious frustration induced ground state
degeneracy.

This kind of frustration can occur either via quenched disorder (i.e.\ 
a random, time-independent distribution of ferromagnetic and
antiferromagnetic spin interactions) or without any disorder, for
example in the fully frustrated antiferromagnetic Ising model on a
triangular lattice. Of course the same problem occurs with XY-spins,
like the XY-antiferromagnet on a triangular or Kagom\'e lattice.  In
this letter we treat exclusively disorder induced frustration.  The
determination of ground states of regularly frustrated systems usually
do not need such algorithmic tools as discussed in this lecture (see
\cite{unfrust} and references therein for a number of examples).

\vfill
\eject

\section{What~is~special~for~simulations~of~disordered~systems?}

As we learned from our simple 4-spin Hamiltonian above, frustration is
often responsible for the existence of many degenerate (or nearly
degenerate) states and metastable states. Suppose one intends to
perform a conventional Monte Carlo simulation with single
spin flip heat bath dynamics of such a system. To explore the whole
energy landscape in order to find the most favorable configurations
one has overcome large energy barriers between the various minima.  As
a consequence, the relaxation times typically become astronomically
large --- not only {\it at} a possible phase transition (in which case
it would be ``critical slowing down''), but also below {\it and}
above. Thus, as is well known in the community of computational
physicists (also among experimentalists, by the way) investigating
disordered or amorphous materials, the {\it equilibration} is nearly
impossible for large systems. Our first commandment in this context 
therefore is

\bc
\fbox{\rule[-7pt]{0cm}{20pt}
{\bf To be modest in system size is mandatory!}}
\ec

Of course everything would be fine if there would be an
efficient\footnote{We emphasize this word, because
it is easy to formulate an algorithm that constructs {\it some}
clusters. Question is, whether the flip acceptance rate is reasonable.}
cluster algorithm at hand, as discussed in this school. However, there
are none --- with a few exceptional cases. To invent an efficient
cluster algorithm for some model, one first has to have a deep
understanding of its physics and a knowledge or an intuition about the
low lying energy configurations, the excitations etc. Thus, {\it cum
grano salis}, if you have understood the system to a rather complete
extent, you might be ready to formulate a cluster algorithm --- with
which you can add some precise numbers (critical exponents etc.) to
your basic understanding. Unfortunately, after several decades of
research we have still not reached this desirable state for most of the 
interesting disordered systems.

The next observation is that different samples, i.e.\ different
disorder realizations, can have completely different dynamical and
static properties. This goes under the name ``large sample to sample
fluctuations'', which originates in the lack of self averaging in some
physical observables. Not all of them show this notorious behavior:
the ground state energy, for instance, is well behaved, simply because
the various local minima are nearly degenerate, but e.g.\ spatial
correlation functions or susceptibilities are not self averaging
quantities. Consider for instance the diluted ferromagnet with site
concentration $c$ and imagine the following two extreme situations: On
a square lattice with $N$ spins one could distribute $N/2$ spins in
such a way that 1) they form a single, compact cluster, or 2) they
occupy a sublattice such that none of them has an occupied nearest
neighbor site. Obviously the magnetization or susceptibility has
completely different characteristics in the two cases: 1) is a bulk
ferromagnet of volume $N/2$ and will have a tendency to order
ferromagnetically at some temperature $T_c$ (in the limit
$N\to\infty$), 2) is a collection of isolated spins that will never
order.

Thus one easily recognizes that the probability distribution
$P_L({\cal O})$ of some observable ${\cal O}$ is usually extremely
broad, in particular non-Gaussian. Rare events (i.e.\ disorder
configurations with small probability) can have a strong impact on
averaged quantities like susceptibilities or autocorrelations.  This
leads to our second commandment for {\it all} investigations of
disordered systems:

\bc\fbox{\rule[-7pt]{0cm}{20pt}
{\bf Sample a huge (!) number of disorder configurations}}
\ec

The study of the probability distribution $P_L({\cal O})$ can be more
useful than only average values.

\section{What can we learn from ground state calculations?}

The ground state is the configuration in which the ``equilibrated''
system settles at exactly zero temperature (if there is more than one,
replace state by states). However, $T=0$ is not accessible in the real
world, so why should we bother? There is a number of reasons for it,
some of them are listed below:

\begin{itemize}

\item[\bf 1)] 
As long as one is interested in {\it equilibrium} properties (and not in
relaxational dynamics, aging, etc.) an {\bf exact} ground state is
more valuable than a non-equilibrium low temperature simulation.

\item[\bf 2)] 
One might expect that some features of the ground state persist at
small temperatures (like the domain structure in the three-dimensional
random field Ising model [fractal or not?], etc.).

\item[\bf 3)] 
If the phase transition into an ordered, may be glassy state, happens
at $T=0$, one can extract critical exponents from ground state
calculations (as for instance in the two-dimensional Ising spin glass).

\item[\bf 4)] 
If the RG (renormalization group) flow for a finite $T$ transition is
governed by a zero-temperature fixed point, one can again extract the
critical exponents via ground state calculations (like in the
three-dimensional random field Ising model). These are then, {\it if}
the RG-picture is correct, identical with those for the finite $T$
transition.

\item[\bf 5)] 
The zero temperature extrapolation of analytical finite $T$
predictions for the glassy phase can be checked explicitly, like in
the SOS (solid-on-solid) model with a disordered substrate.

\end{itemize}

As a motivation this should be enough, in the next section we jump
directly {\it in medias res}. But before we start: many people think
that combinatorial optimization is essentially the traveling salesman
problem, only because it is far the most famous problem (see
\cite{tsp} for an excellent introduction). This is similar to saying
that frustrated disordered systems are essentially spin glasses
(actually the traveling salesman problem and the spin glass problem
are intimately connected via their complexity). One aim of this
lecture is to remove this prejudice and to demonstrate that there are
many more problems out there than only spin glasses (or traveling
salesmen): algorithmically much easier to handle but equally
fascinating. This is also the reason why on the algorithmic side we
focus mainly on network flows: with the help of the material presented
in the appendix everybody should be able to sit down in front of the
computer and to implement efficiently the algorithms discussed there.
If someone wants to know it {\it all}, i.e.\ all background material
on graph theory, linear programming and network flows, we refer to
standard works such as
\cite{wilson_bondy,lawler,papa,chvatal,derigs,ahuja}.

\section{Ground state interface in a random medium}

Although it was historically not the first random Ising model that has
been investigated with the help of the maximum flow / minimum cut
algorithm (this was the random field Ising model, which we shall
discuss later), it might be pedagogically more advantageous to start
with the random bond Ising model with a boundary induced interface.
The reason for the greater intuitive appeal of the latter problem is
that the minimum cut, which the algorithm searches, is identical with
the minimum energy interface of the physical system, which we search.

The random bond Ising ferromagnet (RBIFM) is defined by
\be H=-\sum_{(ij)} J_{ij} \ss_i \ss_j
\label{RBIFM}
\ee 
with $\ss_i=\pm1$ Ising spins and $J_{ij}\ge0$ ferromagnetic
interactions strengths between neighboring spins. These are random
quenched variables, which means that they are distributed according to
some probability distribution and fixed right from the beginning.
$(ij)$ denotes nearest neighbor pairs of a $d+1$-dimensional lattice
of size $L^d\times H$. We denote the coordinates by
$i=(x_1,\ldots,x_d,y)$.

Since the interactions are all ferromagnetic, the ground state is
simply given by $\ss_i=+1$ for all sites $i$ or $\ss_i=-1$ for all sites
$i$. Thus, up to now there is disorder, but no frustration in the
problem. This changes by the boundary conditions (b.c.) we define now:
we apply periodic b.c.\ in the $x$-directions and fix the spins at
$y=1$ to be $+1$ and those at $y=H$ to be $-1$. 
\be
\ss_{(x_1,\ldots,x_d,y=1)}=+1\quad{\rm and}\quad
\ss_{(x_1,\ldots,x_d,y=H)}=-1
\label{bc}
\ee

\begin{figure}
\hfill\epsfxsize=10cm\epsfbox{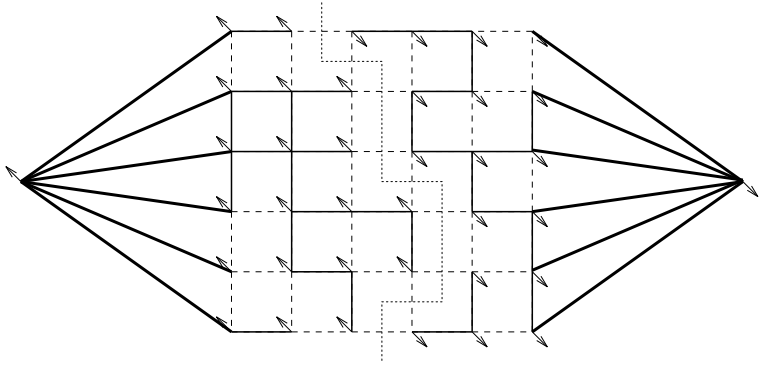}\hfill
\caption{\footnotesize\baselineskip=8pt \label{interface}
  Two-dimensional random bond Ising model with a binary distribution
  of interaction strengths $J_{ij}\in\{J_i,J_2\}$ with $J_1\gg J_2>0$:
  Thick lattice bonds are strong ($J_{ij}=J_1$), broken lattice bonds
  are weak $J_{ij}=J_2$. The left and right boundary spins are
  connected to two extra spins $\ss_\s=+1$ and $\ss_\t=-1$,
  respectively, with infinitely strong bonds.  $\nwarrow$ means
  $\ss_i=+1$, $\searrow$ means $\ss_i=-1$. The dotted line is the
  resulting interface: in this example it passes only weak bonds, by
  which it is of minimal energy. It partitions the lattice into up-and
  down-spins, the bonds (or more precisely: their corresponding
  forward arcs --- see appendix \ref{app_mincut}) that intersect the
  dotted line constitute the set $(S,\overline{S})$
  for the $\s$-$\t$-cut $[S,\overline{S}]$.}
\end{figure}

This induces an interface through the sample where bonds have to be
broken, as indicated in fig.\ \ref{interface}. If all bonds would be
of the same strength $J_{ij}=J$ we would have the pure Ising model and
the interface would simply be a $d$-dimensional hyperplane
perpendicular to the $y$-direction, which costs an energy of $JL^d$,
$J$ for each broken bond. Because of the randomness of the $J_{ij}$ it
is energetically more favorable to break weak bonds: the interface
becomes distorted and its shape is rough.  This model has also been
used to describe fractures in materials where the $J_{ij}$ represents
the local force needed to break the material and it is assumed that
the fracture occurs along the surface of minimum total rupture force.

How do we solve the task of finding the minimal energy configuration
for the interface? First we map it onto a flow problem in a
capacitated network (see appendix A for the nomenclature).  We
introduce two extra sites, a source node $\s$, which we connect to all
spins of the hyperplane $y=1$ with bonds
$J_{\s,(x_1,\ldots,x_d,y=1)}=J_\infty$, and a sink node $\t$, which we
connect to all spins of the hyperplane $y=H$ with bonds
$J_{\s,(x_1,\ldots,x_d,y=H)}=J_\infty$. We choose
$J_\infty=2\sum_{(ij)} J_{ij}$, i.e.\ strong enough that the interface
cannot pass through a bond involving one of the two extra sites.  Now
we enforce the b.c.\ (\ref{bc}) by simply fixing $\ss_\s=+1$ and
$\ss_\t=-1$. The graph underlying the capacitated network we have to
consider is now defined by the set of vertices (or nodes)
\be
N=\{1,\ldots,L^{d+1}\}\cup\{\s,\t\}
\ee
and the set of edges (or arcs) connecting them
\be
A=\{(i,j)|i,j\in N,\; J_{ij}>0\}\;.
\ee
Note that we have forward {\it and} backward arcs for each pair of
interacting sites in the lattice.  The capacities $u_{ij}$ of the arcs
$(i,j)$ is given by the bond strength $J_{ij}$.  For any spin
configuration ${\bf \ss}=(\ss_1,\ldots,\ss_N)$ we define now
\beqn
S            &=&\{i\in N|\ss_i=+1\}\\
\overline{S} &=&\{i\in N|\ss_i=-1\}=N\backslash S
\nonumber
\eeqn
Obviously $\ss_\s\in S$ and $\ss_\t\in\overline{S}$. The knowledge of
$S$ is sufficient to determine the energy of any spin configuration
via (\ref{RBIFM}):
\beqn
H(S) & = &  -\sum_{(i,j)\in E(S)} J_{ij} 
-\sum_{(i,j)\in E(\overline{S})} J_{ij} 
+\sum_{(i,j)\in (S,\overline{S})} J_{ij}
\label{cutenergy}\\
& = & -C + 2\sum_{(i,j)\in (S,\overline{S})} J_{ij} \nonumber
\eeqn
where $E(S)=\{(i,j)|i\in S, j\in S\}$,
$E(\overline{S})=\{(i,j)|i\in\overline{S},j\in\overline{S}\}$ and
$(S,\overline{S})=\{(i,j)|i\in S, j\in \overline{S}\}$.  The constant
$C=\sum_{(i,j)\in E(N)} J_{ij}$ is irrelevant (i.e.\ independent of
$S$). Note that $(S,\overline{S})$ is the set of edges (or arcs)
connecting $S$ with $\overline{S}$, this means it {\bf cuts} $N$ in
two disjoint sets. Since $\s\in S$ and $\t\in \overline{S}$, this is a
so called $\s$-$\t$-cut-set, abbreviated $[S,\overline{S}]$. Thus the
problem of finding the ground state of (\ref{RBIFM}) with the
interface inducing b.c.\ (\ref{bc}) can be reformulated as a {\it
minimum cut} problem
\be
{\rm min}_{S\subset N}\, \{ H'(S) \}
={\rm min}_{[S,\overline{S}]}
\sum_{(i,j)\in(S,\overline{S})} J_{ij}\;.
\label{mincut}
\ee
in the above defined capacitated network (with $H'=(H+C)/2$). It does
not come as a surprise that this minimum cut is {\it identical} with
the interface between the ($\ss_i$=+1)-domain and the
($\ss_i$=-1)-domain that has the lowest energy. Actually any
$\s$-$\t$-cut-set defines such an interface, some of them might
consist of many components, which is of course energetically
unfavorable.

To conclude, we have to find the minimum cut in a capacitated network,
which is, as we show in appendix \ref{app_maxflow}, equivalent to
finding a maximum flow from node $\s$ to node $\t$. An intuitive
argument for this famous max-flow-min-cut theorem is the following:
Suppose you have to push, let's say waterflow through a network of
pipelines, each with some capacity. The capacities in our case are the
ferromagnetic interaction strengths on the bonds (pipes) between the
nodes. Somewhere in the network there is a bottle-neck (in general
consisting of several pipes) which does not allow a further increase
of the waterflow sent from the source to the sink. If the maximum
possible flow goes through the network, the flow on the pipes of the
bottle-neck is equal to their capacity. The minimum cut is simply the
global bottle-neck with the smallest capacity, and thus determines the
maximum flow.

One can solve the above task by applying the straightforward
augmenting path algorithm discussed in appendix \ref{app_aug}, which
is based on the idea to find directed paths in the network on which
one could possibly send more flow from the source to the sink. If one
finds such paths, one augments the flow along them (i.e.\ pushes more
water through the pipes), if there are none, the present flow is
optimal. In the latter case one identifies the corresponding $\s$-$\t$
cut, which then yields the exact ground state interface for the above
problem.

A more efficient way is to use the preflow push algorithm
presented in appendix \ref{app_push}. The idea of this algorithm is to
{\it flood} the network starting from the source. Then one encounters
the situation that some nodes are not able to transport the flood
coming from the source into the direction of the sink, which means
that one has to send some flow back to the source. The time consuming
part of this algorithm is the retreat of floods that have been pushed
too far, and this retreat happens faster if the capacities of backwards
arcs is as large as possible. Bearing this observation in mind
Middleton \cite{middleton} has suggested a nice modification of the
original problem that yields a significant speed up: to forbid {\it
overhangs} of the interface we are discussing is equivalent to
introduce backward arcs with infinite capacity in the corresponding
flow network (obviously a minimum cut will then never contain such an
arc as forward arc). Thus, in the case that too much flow
has been pushed, the retreat works with maximum efficiency.

To conclude let us cite a number of results that have been obtained in
this way. Of particular interest here is the width of the interface
\be
W(L,H)=([y_x^2]_{\rm av}-[y_x]_{\rm av}^2)^{1/2}
\sim L^\zeta\tilde{w}(H/L^\zeta)\;,
\ee
where $y_x$ is the $y$-coordinate of the point $(\underline{x},y)$ of
the interface (note that because of the ``no overhangs'' prescription
$y_x$ is single-valued). $[\ldots]_{\rm av}$ means an average over the
disorder. One expects the finite size scaling form as indicated with a
roughness exponent $\zeta$. From the ground state calculations and
finite size scaling one finds \cite{middleton} that $\zeta=0.41\pm0.01$
in 2d ($L$ up to 120, and $H$ up to 50, with $10^3-10^4$ samples), and
$\zeta=0.22\pm0.01$ in 3d ($L$ up to 30 and $H$ up to 20).

\section{The random field Ising model}

The random field Ising model (RFIM, for a review see \cite{rieger_review}) is
defined
\be
H=-\sum_{(ij)} J_{ij}\ss_i \ss_j - \sum_i h_i\ss_i
\label{RFIM}
\ee
with $\ss_i=\pm1$ Ising spins, ferromagnetic bonds $J_{ij}\ge0$
(random or uniform), $(ij)$ nearest neighbor pairs on a
$d$-dimensional lattice and at each site $i$ a random field $h_i\in R$
that can be positive and negative.  The first term prefers a
ferromagnetic order, which means it tries to align all spins. The
random field, however, tends to align the spins with the field which
points in random directions depending on whether it is positive or
negative. This is the source of frustration in this model.

Let us suppose for the moment uniform interactions $J_{ij}=J$ and a
symmetric distribution of the random fields with mean zero and
variance $h_r$.  It is established by now that in 3d (and higher
dimensions) the RFIM shows ferromagnetic long range order at low
temperatures, provided $h_r$ is small enough. In 1d and 2d there is no
ordered phase at any finite temperature. Thus in 3d one has a
paramagnetic/ferromagnetic phase transition along a line $h_c(T)$
in the $h_r$-$T$-diagram.

The renormalization group (RG) picture says that the nature of the
transition is the same\footnote{We leave aside the discussion
about a possible tricritical point (which does not seem to be the
case \protect{\cite{rieger_review}}) and the existence of an intervening spin
glass phase.} all along the line $h_c(T)$, with the exception being the
pure fixed point at $h_r=0$ and $T_c\sim4.51J$. The RG flow is
dominated by a zero temperature fixed point at $h_c(T$=$0)$. As
a consequence, the critical exponents determining the critical behavior
of the RFIM should be all identical along the phase transition line,
in particular identical to those one obtains {\it at zero temperature}
by varying $h_r$ alone. 

Therefore we consider zero temperature from now on. Close to the
transition at $h_c=h_c(T$=$0)$ one would e.g.\ expect for the
disconnected susceptibility
\be
\chi_{\rm dis}=\frac{1}{L^3}\biggl[\sum_{i,j}\ss_i\ss_j\biggr]_{\rm av}
\sim L^{4-\overline{\eta}}\tilde{\chi}(L^{1/\nu}\delta)
\ee
where $\delta=h_r-h_c$ is the distance from the critical point
and $\nu$ is the thermal critical exponent. An analogous expression
holds for the magnetization involving the exponent $\beta$. Thus to
estimate a set of critical exponents the task is to calculate the
ground state configurations of the RFIM (\ref{RFIM}).

This optimization task is again equivalent to a maximum flow problem
\cite{barahona_rfim}, as in the interface model discussed in the last
section. Historically the RFIM was the first physical model that has
been investigated with a maximum flow
algorithm \cite{ogielski}. However, here the minimum-cut is not a
geometric object within the original system and therefore we found it
more intuitive to discuss the RFIM after the interface problem.

In essence we proceed in the same way as in the last section. Again we
add to extra nodes $\s$ and $\t$ and put spins with fixed values
there:
\be
\ss_\s=+1\quad{\rm and}\qquad \ss_\t=-1
\ee
We connect all sites with positive random field to the node $\s$ and
all sites with negative random field to $\t$:
\be
\ba{lcl}
J_{\s i} & = & \left\{
\ba{lcl} 
h_i & \quad &{\rm if}\; h_i\ge0\\
0 & \quad &{\rm if}\; h_i<0
\ea\right.
\\
J_{i\t} & = & \left\{
\ba{lcl} 
|h_i| & \quad &{\rm if}\; h_i<0\\
0 & \quad &{\rm if}\; h_i\ge0
\ea\right.
\ea
\ee
We construct a network with the set of nodes
$N=\{1,\cdots,L^d\}\cup\{\s,\t\}$ and the set of (forward and
backward) arcs $A=\{(i,j)|\,i,j\in N,\;J_{ij}>0\}$. Each of them has a
capacity $u_{ij}=J_{ij}$. Now we can write the energy or cost
function as
\be
E=-\sum_{(i,j)\in A} J_{ij} \ss_i \ss_j
\ee
and, by denoting the set $S=\{i\in N|S_i=+1\}$ and
$\overline{S}=N\backslash S$ the energy can be written as in equation
(\ref{cutenergy}):
\be
E(S)=-C+2\sum_{(i,j)\in(S,\overline{S})} J_{ij}
\ee
with $C=\sum_{(i,j)\in A} J_{ij}$. The problem is reduced to the problem
of finding a minimum $\s$-$\t$-cut as in (\ref{mincut}). The difference
to the interface problem is that now the extra bonds connecting the
two special nodes $\s$ and $\t$ with the original lattice do not have
infinite capacity: they can lie {\it in} the cut, namely whenever it is more
favorable not to break a ferromagnetic bond but to disalign a spin with
its local random field. In the extended graph which we consider the
$\s$-$\t$-cut again forms connected interface, however, in the
original lattice (without the bonds leading to and from the extra
nodes) the resulting structure is generally {\it disconnected}, a
multicomponent interface. Each single component surrounds a connected
region in the original lattice containing spins, which all point in
the same direction. In other words, they form ferromagnetically
ordered domains separated by domain walls given by the subset of the
$\s$-$\t$-cut that lies in the original lattice. 

The maximum flow algorithm has been used by Ogielski \cite{ogielski} to
calculate the critical exponents of the RFIM via the above mentioned
finite size scaling. He obtained
\be
\nu=1.0\pm0.1\;,\qquad\overline{\eta}=1.1\pm0.2\;,\qquad\beta\approx0.05
\ee
with $\beta$ being so small that it is (numerically) indistinguishable
from zero, indicating a {\it discontinuous} transition. These
estimates are compatible with those obtained by recent Monte Carlo
simulations supporting the RG idea of the universality of the
transition at finite {\it and} zero temperature. However, this is
still not the end of the story, since various scaling predictions,
also based on the RG picture, are violated. For further details we
refer to the review \cite{rieger_review}.

\section{The diluted antiferromagnet in an external field}

Experimentally it is of course hard to prepare a random field at each
lattice site, therefore one might ask why people have been so
enthusiastic about the RFIM, discussed in the last section, over
decades.  Actually it is because within a field theoretic perturbation
theory (around small random fields) it has been shown \cite{fishman}
that the RFIM is in the same universality class as the diluted
antiferromagnet in a {\it uniform} magnetic field (DAFF) defined via
\be
H=+\sum_{(ij)} J_{ij}\varepsilon_i\varepsilon_j\ss_i\ss_j
-\sum_i h_i\varepsilon_i\ss_i
\label{DAFF}
\ee
where $\ss_i=\pm1$, $J_{ij}\ge0$, $(ij)$ are nearest neighbor pairs on
a simple cubic lattice, and $\varepsilon_i\in\{0,1\}$ with
$\varepsilon_i=1$ with probability $c$, representing the concentration
of spins. Usually one takes $J_{ij}=J$ and $h_i=h$, both uniform.
Because of the plus sign in front of the first term in (\ref{DAFF})
all interactions are antiferromagnetic, the model represents a diluted
antiferromagnet, for which many experimental realizations exist (e.g.\
Fe$_c$Zn$_{1-c}$F$_2$).  Now that neighboring spins tend to point in
opposite directions due to their antiferromagnetic interaction a
uniform field competes with this ordering tendency by trying to align
them all. Thus it is again a frustrated system. Due to the analogy to
the RFIM model one expects at low temperatures and small enough fields
a second order phase transition from a paramagnetic to an
antiferromagnetic phase.

In recent years people began to doubt the folklore that the DAFF is
under all circumstances a good experimental realization of the RFIM
model. Note that this result has been derived for small fields $h$,
and the question is whether this still holds at larger fields. The
largest field value at which the paramagnetic-antiferromagnetic
transition can be studied is $h_c(T=0)$. This motivates the study of
the ground state transition along the same lines as in the RFIM
context. Preliminary results \cite{esser} indicate that the critical
exponents are {\it different} here, which implies that the RFIM and
the DAFF are in different universality classes at large field values.

Here we are primarily interested in the question whether we can again
map the calculation of ground states onto a maximum flow problem, as
for the RFIM. The answer is yes as long as the antiferromagnetic
interactions are short ranged among nearest neighbors on a bipartite
lattice.  With zero external field the ground state would be
antiferromagnetic, which means (remember we have a simple cubic
lattice) that we can define two bipartite sublattices $A$ and $B$ like
the black and white fields of a checkerboard. Each site $i$ in $A$
finds all its nearest neighbors $j$ in $B$ and vice versa. Define new
spin and field variables via
\be
\ss_i'=\left\{
\ba{lcl}
+\ss_i & \quad{\rm for}\quad & i\in A\\
-\ss_i & \quad{\rm for}\quad & i\in B
\ea\right.
\quad,\quad
h_i'=\left\{
\ba{lcl}
+\varepsilon_i h_i & \quad{\rm for}  \quad & i\in A\\
-\varepsilon_i h_i & \quad{\rm for}\quad & i\in B
\ea\right.\;.
\nonumber
\ee
Since $\ss_i'\ss_j'=-\ss_i\ss_j$ for all
nearest neighbor pairs $(ij)$ one can write (\ref{DAFF}) as
\be
H=-\sum_{(ij)} J_{ij}' \ss_i'\ss_j'
-\sum_i h_i' \ss_i'
\ee
with $J_{ij}'=J_{ij}\varepsilon_i\varepsilon_j$. Now the Hamiltonian
has exactly the same form as the one for the RFIM, since
$J'_{ij}\ge0$.  Note that even if one starts with uniform bonds
$J_{ij}=J$ and a uniform field $h_i=h$ the dilution generates bond-
and field disorder!  Now that one has reduced the problem to the RFIM
we also know how to map it to a maximum flow problem. Hartmann and
Usadel \cite{hartmann} have extended the algorithm in such a way that
{\it all} ground states can be calculated: for uniform bonds and fields
the resulting RFIM has a discrete distribution of random bonds and
fields, which leads in general to a high degeneracy of the
ground state, something that does not happen in case of a uniform
distribution, where usually the ground state is unique.

In this context we would like to
mention the Coulomb glass model \cite{shklovsky,tenelsen}, which is a
model for point charges on a $d$-dimensional lattice with long-range
Coulomb interactions (repulsive of strength $V/r$ with $V$ {\it
positive} and $r$ being the Euclidean distance between two charges):
\be
H=\sum_{i,j} \frac{V}{r_{ij}} n_i n_j + \sum_i n_i\mu_i
\ee
where now the sum is over {\it all} pairs of sites of the lattice.
$n_i\in\{0,1\}$ indicates the presence ($n_i=1$) or absence ($n_i=0$)
of a charged particle at site $i$ and $r_{ij}$ is the Euclidean
distance between site $i$ and site $j$. The random local potentials
$\mu_i\in[-W,W]$ represent the quenched disorder. Obviously this model
is equivalent to an antiferromagnet with long-range interactions and
random fields.  Up to now no way of mapping this interesting problem
onto a network flow problem is known, it seems to fall into the spin
glass class, which we discuss now.

\section{The spin glass problem}

Spin glasses are the prototypes of (disordered) frustrated systems,
their history is quite a long one and for the present status of
numerical investigation I refer to \cite{rieger_review}, where also
numerous references to experimental and theoretical introductions can
be found.  In the models we discussed up to now, the frustration was
caused by two separate terms of different physical origin
(interactions and external fields or boundary conditions). Spin
glasses are magnetic systems in which the magnetic moments interact
ferro- or antiferromagnetically in a random way, as in the following
Edwards-Anderson Hamiltonian for a short ranged Ising spin glass (SG)
\be
H=-\sum_{(ij)} J_{ij} \ss_i \ss_j\;,
\label{SG}
\ee
where $\ss_i=\pm1$, $(ij)$ are nearest neighbor interactions on a
$d$-dimensional lattice and the interaction strengths $J_{ij}\in R$
are unrestricted in sign. In analogy to eq.\
(\ref{cutenergy}--\ref{mincut}) one shows that the problem of finding
the ground state is again equivalent to finding a minimal cut
$[S,\overline{S}]$ in a network
\be
{\rm min}_{\bf\underline{\ss}}\, \{ H'({\bf\underline{\ss}}) \}
={\rm min}_{[S,\overline{S}]}
\sum_{(i,j)\in(S,\overline{S})} J_{ij}\;,
\label{sgcut}
\ee
again $H'=(H+C)/2$ with $C=\sum_{(ij)}J_{ij}$.
However, now the capacities $u_{ij}=J_{ij}$ of the underlying network
are {\it not} non-negative any more, therefore it is {\it not} a
minimum-cut problem in the sense of appendix \ref{app_mincut} and
thus it is also not equivalent to a maximum flow problem, which we
know how to handle efficiently.

It turns out that the spin glass problem is {\it much} harder than the
questions we have discussed so far. In general (i.e.\ in any dimension
larger than two and also for 2d in the presence of an external field)
the problem of finding the SG ground state is ${\cal NP}$-complete
\cite{barahona}, which means in essence that no polynomial algorithm
for it is known (and also that chances to find one in the future are
marginal). Nevertheless, some extremely efficient algorithms for it
have been developed \cite{groetschel,juenger,kobe}, which have a
non-polynomial bound for their worst case running-time but which
terminate (i.e\ find the optimal solution) after a reasonable
computing time for pretty respectable system sizes.

\begin{figure}
\hfill\epsfxsize=6cm\epsfbox{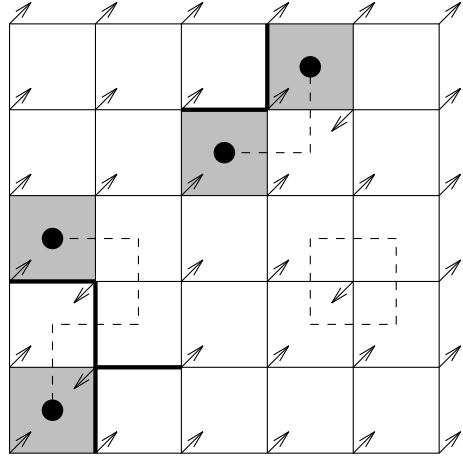}\hfill
\caption{\footnotesize\baselineskip=8pt \label{frust} Two-dimensional
  Ising spin glass with $\pm$-J couplings: Thin lines, are positive
  interactions, thick lines are negative interactions, $\nearrow$
  means $\ss_i=+1$, $\swarrow$ means $\ss_i=-1$, shaded faces are
  frustrated plaquettes, broken lines cross unsatisfied edges.
}
\end{figure}

First we discuss the only non-trivial case that can be solved with a
polynomial algorithm: the two-dimensional Ising SG on a planar graph.
This problem can be shown to be equivalent to finding a minimum weight
perfect matching, which can be solved in polynomial time. We do not
treat matching problems and the algorithms to solve them in this
lecture (see \cite{lawler,papa,derigs}), however, we would like to
present the idea \cite{barahona}. To be concrete let us consider a
square lattice with free boundary conditions. Given a spin
configuration ${\bf \underline{\ss}}$ (which is equivalent to $-{\bf
\underline{\ss}}$) we say that an edge (or arc) $(i,j)$ is satisfied
if $J_{ij}\ss_i\ss_j>0$ and it is {\it unsatisfied} if
$J_{ij}\ss_i\ss_j<0$.  Furthermore we say a plaquette (i.e.\ a unit
cell of the square lattice) is {\it frustrated} if it is surrounded by
an odd number of negative bonds (i.e.\ $J_{ij}\cdot J_{jk}\cdot
J_{kl}\cdot J_{li} <0$ with $i$, $j$, $k$ and $l$ the four corners of
the plaquette)). There is a one-to-one correspondence between
equivalent spin configurations (${\bf \underline{\ss}}$ and $-{\bf
\underline{\ss}}$) and sets of unsatisfied edges with the property
that on each frustrated (unfrustrated) plaquette there is an odd
(even) number of unsatisfied edges. See fig.\ \ref{frust} for
illustration.

Note that
\be
H({\bf\underline{\ss}})=-C+2\sum_{\rm unsatisfied\;edges} |J_{ij}|\;.
\ee
which means that one has to minimize the sum over the weights of
unsatisfied edges. A set of unsatisfied edges will be constituted by a
set of paths (in the dual lattice) from one frustrated plaquette to
another and a set of closed circles (see fig.\ \ref{frust}).
Obviously the latter always increase the energy so that we can neglect
them. The problem of finding the ground state is therefore equivalent
to finding the minimum possible sum of the weights of these paths
between the frustrated plaquettes.  This means that we have to {\it
match} the black dots in the fig.\ \ref{frust} with one another in an
optimal way. One can map this problem on a minimum weight {\it perfect
matching}\footnote{A perfect matching of a graph $G=(N,A)$ is a set
$M\subseteq A$ such that each node has only has only one edge of $M$
adjacent to it.} problem, which can be solved in polynomial time (see
\cite{barahona} for further details).

Note that for binary couplings, i.e.\ $J_{ij}=\pm J$, where
$J_{ij}=+J$ with probability $p$ the weight of a matching is simply
proportional to the sum of the lengths of the various paths connecting
the centers of the frustrated plaquettes, which simplifies the actual
implementation of the algorithm. In \cite{kawashima} the 2d $\pm J$
spin glass and the site disordered SG\footnote{The site disordered
spin glass is defined as follows: occupy the sites of a square lattice
randomly with $A$ (with concentration $c$) and $B$ (with concentration
$1-c$) atoms. Now define the interactions $J_{ij}$ between neighboring
atoms: $J_{ij}=-J$ if on both sites are $A$-atoms and $J_{ij}$
otherwise.} has been studied extensively with this algorithm.

As we mentioned, in any other case except the planar lattice situation
discussed above the spin glass problem is $\cal NP$-hard. In what
follows we would like to sketch the idea of an efficient but
non-polynomial algorithm \cite{juenger,diehl}. To avoid confusion with the
minimum cut problem we discussed in connection with maximum flows one
calls the problem (\ref{sgcut}) a max-cut problem (since finding the
minimum of $H$ is equivalent to finding the maximum of $-H$). 

Let us consider the vector space $R^A$. For each cut
$[S,\overline{S}]$ define $\chi^{(S,\overline{S})}\in R^A$, the
incidence vector of the cut, by $\chi_e^{(S,\overline{S})}=1$ for each
edge $e=(i,j)\in(S,\overline{S})$ and $\chi_e^{(S,\overline{S})}=0$
otherwise. Thus there is a one-to-one correspondence between cuts in
$G$ and their $\{0,1\}$-incidence vectors in $R^A$. The {\it
cut-polytope} $P_C(G)$ of $G$ is the convex hull of all incidence
vectors of cuts in $G$: $P_C(G)={\rm
conv}\;\{\chi^{(S,\overline{S})}\in R^A\,|\,S\subseteq A \}$.  Then
the max-cut problem can be written as a {\it linear program}
\be
{\rm max}\,\{\underline{u}^T\underline{x}\,|\,\underline{x}\in P_C(G)\}
\ee
since the vertices of $P_C(G)$ are cuts of $G$ and vice versa.  Linear
programms usually consist of a linear cost function
$\underline{u}^T\underline{x}$ that has to be maximized under the
constraint of various inequalities defining a Polytope in $R^n$ (i.e.\
the convex hull of finite subsets of $R^n$) and can be solved for
example by the simplex method, which proceeds from corner to corner of
that polytop to find the maximum (see e.g.\
\cite{lawler,chvatal,derigs}). The crucial problem in the present case
is that it is $\cal NP$-hard to write down all inequalities that
represent the cut polytop $P_C(G)$.

It turns out that also {\it partial} systems are useful, and this is
the essential idea for an efficient algorithm to solve the general
spin glass problem as well as the traveling salesman problem or other
so called mixed integer problems (i.e.\ linear programms where some of
the variables $x$ are only allowed to take on some integer values,
like 0 and 1 in our case) \cite{tsp,thienel}. One chooses a system of
linear inequalities $L$ whose solution set $P(L)$ contains $P_C(G)$
and for which $P_C(G)={\rm convex\;hull}\,\{{\bf x}\in P(L)|x\,\,{\rm
  integer}\}$. In the present case these are $0\le x\le 1$, which is
trivial, and the so called cycle inequalities, which are based on the
observation that all cycles in $G$ have to intersect a cut an even
number of times (have a look at the cut in fig.\ 1 and choose as
cycles for instance the paths around elementary plaquettes). The most
remarkable feature of this set $L$ of inequalities is that the
separation problem\footnote{The seperation problem for a set of
  inequalities $L$ consists in either proving that a vector $x$
  satisfies all inequlaities of this class or to find an inequality
  that is violated by ${\bf x}$. A linear programm can be solved in
  polynomial time if and only if the separation problem is solvable in
  polynomial time \protect{\cite{groetschel2}}.} for them can be solved
in polynomial time: the {\bf cutting plane algorithm} which,
starting from some small initial set of inequalities, generates
iteratively new inequalities until the optimal solution for the actual
subset of inequalities is feasible. Note that one does not solve this
linear programm by the simplex method since the cycle inequalities are
still too numerous for this to work efficiently.

Due to the insufficient knowledge of the inequalities that are
necessary to describe $P_C(G)$ completely, one may end up with a
nonintegral solution ${\bf x}^*$. In this case one {\bf branches} on some
fractional variable $x_e$ (i.e.\ a variable with
$x_e^*/\!\!\!\!\!\in\{0,1\}$), creating two subproblems in one of which
$x_e$ is set to 0 and in the other $x_e$ is set to 1. Then one applies
the cutting plane algorithm recursivley for both subproblems, which is
the origin of the name {\bf branch-and-cut}. Note that in principle
this algorithm is not restricted to any dimension, boundary conditions,
or to the fieldless case. However, there are realizations of it that run
fast (e.g.\ in 2d) and others that run slow (e.g.\ in 3d) and it is
ongoing research to improve on the latter. A detailed description 
of the rather complex algorithm can be found in \cite{thienel,diehl}.

The typical questions one tries to address in the context of spin
glasses is: is there a spin glass transition at finite temperature,
below which the spins freeze into some configuration (i.e.\ $\langle
\ss_i\rangle_T\ne 0$ for $T<T_c$). What can we do with ground state
calculation to answer this question? Here the concept of the domain
wall energy plays a crucial role \cite{bray}. What a finite but small
temperature does is to destroy the ground state order by
collectively flipping larger and larger clusters (droplets). If the
energy cost for a reversal of a cluster of linear size $L$ increases
with $L$ (like $\Delta E\propto L^y$ with $y>0$) thermal fluctuation
will not be able to destroy long range order, and thus we have a spin
glass transition at finite $T_c$. If it decreases (i.e.\ $y<0$) long
range order is unstable with respect to thermal fluctuations and the
spin glass state occurs only at $T=0$. As an example consider the
$d$-dimensional {\it pure} Ising ferromagnet, for which the
ground state is all spins up or all down. Reversing a cluster of linear
size $L$ breaks all surface bonds of this cluster, which means that
it costs an energy $\Delta E\propto L^{d-1}$, i.e.\ $y=d-1$ for the
pure ferromagnet. Thus the ferromagnetic state in pure Ising systems
is stable for $d>1$, which is well known.  Instead of reversing spins
one usually studies the energy difference between ground states for
periodic and antiperiodic boundary conditions. In \cite{rieger_sg} it
has been shown that 
\be
\Delta E\sim L^y
\ee
with $y=-0.281$ for the 2d Ising spin glass with a uniform
distribution (thus there is no finite $T$ SG transition in this
case). It has been speculated that in the $\pm J$ case for a range of
concentration of ferromagnetic bonds \cite{antiphase} and in the
site-random case for some concentration of $A$ atoms \cite{siterandom}
a spin glass phase might exist at non-zero temperature $T>0$. This
possibility has been ruled out in \cite{kawashima} with the help of
ground state calculations.

With the above mentioned branch \& cut algorithm the magnetic field
dependence of the ground state magnetization $m_L(h)=L^{-d}[|\sum_i
\ss_i|]_{\rm av}$ has been calculated in the 2d case with a uniform
coupling distribution. In \cite{rieger_sg} it has been shown that it
obeys finite size scaling form
\be
m_L(h)\sim L^{-d/2}\tilde{m}(Lh^{1/\delta})
\ee
(note $d=2$) with $\delta=1.481$. This value is remarkable in so far
as it clearly violates the scaling prediction $\delta=1-y$.

Finally we would like to focus our attention on the very important
concept of {\it chaos} in spin glasses. This notion implies an extreme
sensitivity of the SG-state with respect to small parameter changes
like temperature or field variations. There is a length scale
$\lambda$ --- the so called overlap length --- beyond which the spin
configurations within the same sample become completely decorrelated
if compared for instance at two different temperatures
\be
C_{\Delta T}:=
[\langle \ss_i \ss_{i+r}\rangle_T 
\langle \ss_i \ss_{i+r}\rangle_{T+\Delta T}]_{\rm av}
\sim \exp\Bigl(-r/\lambda(\Delta T)\Bigr)\;.
\ee
This should also hold for the ground states if one slightly varies the
interaction strengths $J_{ij}$ in a random manner with amplitude
$\delta$. Let ${\bf\underline{\ss}}$ be the ground state of a sample
with couplings $J_{ij}$ the ground state of a sample with couplings
$J_{ij}+\delta K_{ij}$, where the $K_{ij}$ are random (with zero mean
and variance one) and $\delta$ is a small amplitude. Now define
the overlap correlation function as
\be
C_\delta(r):=[\ss_i\ss_{i+r}\,\ss_i'\ss_{i+r}']_{\rm av}
\;\sim\; \tilde{c}(r\delta^{1/\zeta})\;,
\ee
where the last relation indicates the scaling behavior we would expect
(the overlap length being $\lambda\sim\delta^{-1/\zeta}$) and $\zeta$
is the {\it chaos} exponent. In \cite{rieger_sg} this scaling
prediction was confirmed with $1/\zeta=1.2\pm0.1$.

\section{The SOS-model on a disordered substrate}
\label{sos_section}

\begin{figure}
\hfill\epsfxsize=9cm\epsfbox{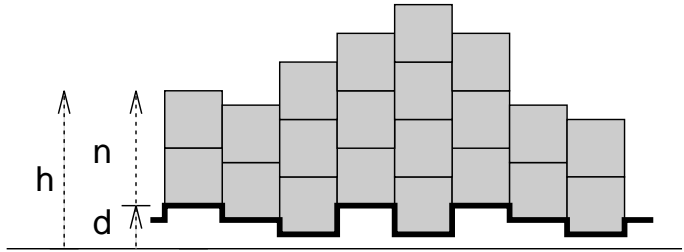}\hfill
\caption{\footnotesize\baselineskip=8pt \label{sos} The SOS model on a
  disordered substrate. The substrate heights are denoted by
  $d_i\in[0,1]$, the number of particle on site $i$ by $n_i\in Z$,
  which means that they could also be negative, and the total height
  on site $i$ by $h_i=d_i+n_i$ }
\end{figure}

Up to now we have considered Ising models exclusively. Quite recently
it has been shown \cite{blasum,rieger_sos} that many more frustrated
systems are amenable to ground state studies of the kind we discussed
so far. Consider a solid-on-solid model with random offsets, modeling
a crystalline surface on a disordered substrate as indicated in fig.\ 
\ref{sos}. It is defined by the following Hamiltonian (or energy
function):
\be H=\sum_{(ij)} f(h_i-h_j)
\label{sos_ham}
\ee
where $(ij)$ are nearest neighbor pairs on a
$d$--dimensional lattice ($d=1,2$) and $f(x)$ is an arbitrary convex
($f''(x)\ge0$) and symmetric ($f(x)=f(-x)$) function, for instance
$f(x)=x^2$. Each height variable $h_i=d_i+n_i$ is the sum of an
integer particle number which can also be negative, and a substrate
offset $d_i\in[0,1[$. For a flat substrate, $d_i=0$ for all sites $i$,
we have the well known SOS-model \cite{puresos}. The disordered
substrate is modeled by random offsets $d_i\in[0,1[$ \cite{tsai},
which are distributed independently.

The model (\ref{sos_ham}) has a phase transition at a temperature
$T_c$ from a (thermally) rough phase for $T>T_c$ to a {\it super-rough}
low temperature phase for $T<T_c$. In two dimension "rough" means that
the height-height correlation function diverges logarithmically with
the distance $C(r)=[\langle(h_i-h_{i+r})^2\rangle]_{\rm av}=a\cdot
T\cdot \log(r)$ (with $a=1/\pi$ for $f(x)=x^2$), "super-rough" means
that either the prefactor on front of the logarithm is significantly
larger than $a\cdot T$, or that $C(r)$ diverges faster than $\log(r)$,
e.g.\ $C(r)\propto\log^2(r)$.

A part of the motivation to study this model thus comes from its
relation to flux lines in disordered superconductors, in particular
high-T$_c$ superconductors: The phase transition occurring for
(\ref{sos_ham}) is in the same universality class as a flux
line array with point disorder defined via the two-dimensional
Sine-Gordon model with random phase shifts
\be
H=-\sum_{(ij)} (\phi_i-\phi_j)^2
-\lambda\sum_i\cos(\phi_i-\theta_i)\:,
\label{sine_ham}
\ee
where $\phi_i\in[0,2\pi[$ are phase variables, $\theta_i\in[0,2\pi[$
are quenched random phase shifts and $\lambda$ is a coupling constant.
One might anticipate that both models (\ref{sos_ham}) and
(\ref{sine_ham}) are closely related by realizing that both have the
same symmetries (the energy is invariant under the replacement $n_i\to
n_i+m$ ($\phi_i\to\phi_i+2\pi m$) with $m$ an integer). Close to the
transition one can show that all higher order harmonics apart from the
one present in the Sine-Gordon model (\ref{sine_ham}) are irrelevant
in a field theory for (\ref{sos_ham}), which establishes the identity
of the universality classes\footnote{Note, however, that that far away
  from $T_c$, as for instance at zero temperature, there might be
  differences in the two models.}.

To calculate the ground states of the SOS model on a disordered
substrate with general interaction function $f(x)$ we map it onto a
minimum cost flow model. Let us remark, however, that the special case
$f(x)=|x|$ can be mapped onto the interface problem in the random bond
Ising ferromagnet in 3d with {\it columnar} disorder \cite{zeng}
(i.e.\ all bonds in a particular direction are identical), by which it
can be treated with the maximum flow algorithm we know already. 

\begin{figure}
\hfill\epsfxsize=11cm\epsfbox{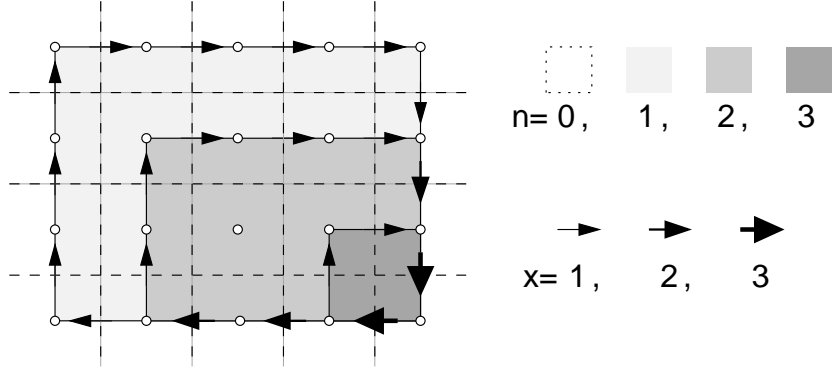}\hfill
\caption{\footnotesize\baselineskip=8pt \label{cycles} 
  The flow representation of a surface (here a "mountain" of height
  $n_i=3$).  The broken lines represent
  the original lattice, the open dots are the nodes of the dual
  lattice. The arrows indicate a flow on the dual lattice, that
  results from the height differences of the variables $n_i$ on the
  original lattice. Thin arrows indicate a height difference of
  $x_{ij}=1$, medium arrows $x_{ij}=2$ and thick arrows $x_{ij}=3$.
  According to our convention the larger height values are always on
  the {\it right} of an arrow. Observe that on each node one the mass
  balance constraint (\protect{\ref{constraint}}) is fulfilled.}
\end{figure}

We define a network $G$ by the set of nodes $N$ being the sites of the
dual lattice of our original problem and the set of {\it directed}
arcs $A$ connecting nearest neighbor sites (in the dual lattice)
$(i,j)$ and $(j,i)$. If we have a set of height variables $n_i$ we
define a flow ${\bf x}$ in the following way: Suppose two neighboring
sites $i$ and $j$ have a positive (!)  height difference $n_i-n_j>0$.
Then we assign the flow value $x_{ij}=n_i-n_j$ to the directed arc
$(i,j)$ in the dual lattice, for which the site $i$ with the larger
height value is on the right hand side, and assign zero to the
opposite arc $(j,i)$, i.e.\ $x_{ji}=0$. And also $x_{ij}=0$ whenever
site $i$ and $j$ are of the same height. See fig.\ \ref{cycles} for a
visualization of this scheme. Obviously then we have:
\be
\forall i\in N\;:\qquad
\sum_{\{ j\,|\,(i,j)\in A \}} x_{ij} = 
\sum_{\{ j\,|\,(j,i)\in A \}} x_{ji}\;.
\label{constraint}
\ee
On the other hand, for an arbitrary set of values for $x_{ij}$ the
constraint (\ref{constraint}) has to be fulfilled in order to be a
flow, i.e.\ in order to allow a reconstruction of height variables out
from the height differences. This observation becomes immediately
clear by looking at fig.\ \ref{cycles}.  

We can rewrite the energy
function as
\be
H({\bf x})=\sum_{(i,j)} h_{ij}(x_{ij})\;,\qquad{\rm with}\qquad
c_{ij}(x)=f(x-d_{ij})\;,
\ee
with $d_{ij}=d_i-d_j$. Thus our task is to minimize $H({\bf x})$ under
the constraint (\ref{constraint}), which is (see appendix
\ref{app_convex}) a {\it minimum-cost-flow} problem with the mass
balance constraints (\ref{constraint}) and arc {\it convex} cost
functions $h_{ij}(x_{ij})$. It is worth mentioning that this mapping
from the SOS model to a flow problem is closely related to the dual
link representation of the XY-model in two dimensions
\cite{kleinert}. This does not come as a great surprise since we already
pointed out the relationship with the Sine-Gordon Hamiltonian
involving phase variables (\ref{sine_ham}).

The most straightforward way to solve this problem is to start with
all height variables set to zero (i.e.\ ${\bf x}=0$) and then to look
for regions (or clusters) that can be increased collectively by one
unit with a gain in energy. This is essentially what the negative
cycle canceling algorithm discussed in appendix \ref{app_convex} does:
The negative cycles in the dual lattice surround the regions in which
the height variables should be increased or decreased by one. However,
it turns out that this is a non-polynomial algorithm, the so called
successive shortest path algorithm is more efficient and solves this
problem in polynomial time, see appendix \ref{app_convex}. This
algorithm starts with an optimal solution for $H({\bf x})$, which is
given by $x_{ij}=+1$ for $d_{ij}>1/2$, $x_{ij}=-1$ for $d_{ij}<-1/2$
and $x_{ij}=0$ for $d_{ij}\in[-1/2,+1/2]$. Since this set of flow
variables violates the mass balance constraints (\ref{constraint})
(in general there is some imbalance at the nodes) the algorithm
iteratively removes the excess/deficit at the nodes by augmenting
flow.

Let us briefly summarize the results one obtains by applying this
algorithm to the ground state problem for the SOS model on a disordered
substrate \cite{rieger_sos}:

\begin{itemize}
\item The height-height correlation function diverges like
$C(r)\propto\log^2(r)$ with the distance $r$.
\item $\chi_L=L^{-4}\sum_{i,j} [(h_i-h_j)^2]_{\rm av}$ can nicely be
fitted to $\chi_L=a+b\log(L)+c\log^2(L)$, indicating again a $\log^2$
dependence of the height-height correlation function. Moreover, the
coefficients $a$, $b$ and $c$ depend on the power $n$ in $f(x)=|x|^n$:
$c$ increases systematically with increasing $n$.
\item By considering a boundary induced step in the ground state
configuration one sees that the step energy increases algebraically
with the system size: $\Delta E\sim L^y$ with $y=0.45\pm0.05$.  This
corresponds to the domain wall energy introduced in the context of
spin glasses in the las section.  Furthermore the step is {\it
fractal} with a fractal dimension close to $3/2$.
\item Upon a small, random variation of the substrate heights $d_i$ of
amplitude $\delta$ the ground state configuration decorrelates beyond
a length scale $\lambda\sim\delta^\eta$ with $\eta=0.95\pm0.05$. This
implies the {\it chaotic} nature of the glassy phase in this model in
analogy to spin glasses.
\end{itemize}

We would like to mention that this mapping of the original SOS model
(\ref{sos_ham}) on the flow problem works only for a planar graph
(i.e.\ free or fixed boundary conditions), otherwise it is not always
possible to reconstruct the height variables $n_i$ from the height
differences $x_{ij}$. As a counterexample in a toroidal topology
(periodic boundary conditions) consider a flow, which is zero
everywhere except on a circle looping the torus, where it is one.
Although this flow fulfills the mass balance constraints (which are
local) it is globally inadmissible: To the right of this circle the
heights should be one unit larger than on left, but left and right
become interchanged by looping the torus in the perpendicular
direction, which causes a contradiction. If one insists on periodic
boundary conditions, which have some advantages due to the
translational invariance, one should recur to the special case
$f(x)=|x|$, which can be treated differently, as we mentioned in the
beginning of this section.

\section{Vortex glasses and traffic}

Finally we would like to focus on some further applications of the
minimum cost flow algorithms that we discussed in the last section.
Since we deal with network flow problems it should not come as a
surprise, that a number of physical problems involving magnetic flux
lines can be mapped onto them. We already mentioned the Sine-Gordon
model with random phase shifts (\ref{sine_ham}) describing a flux line
array with point disorder and which is related to the SOS model on a
disordered substrate. This relationship can be made more concrete with
the help of the triangular Ising SOS model as discussed in
\cite{zeng}.

The gauge glass model describes the {\bf vortex glass} transition in
three-dimensional superconductors. If one includes the screening of the
interactions between vortices one can show that in the strong
screening limit, the model Hamiltonian (in the link representation)
acquires the form \cite{wengel}
\be
H=\sum_{(i,j)} (x_{ij}-b_{ij})^2
\label{vortex}
\ee
where $x_{ij}$ are integer vortex variables living on the links
$(i,j)$ of the dual of the original simple cubic lattice. They
represent magnetic flux lines, by which they have to be divergenceless
--- which means that they have to fulfill the mass balance constraint
(\ref{constraint}). The quenched random variables $b_{ij}$ also
fulfill the same constraint (they have to be constructed as a lattice
curl from a quenched vector potential). Moreover one has periodic
boundary condition.

It has been shown that this model has a vortex glass transition at
zero temperature. Thus, for the characterization of the critical
behavior either low temperature Monte Carlo simulations \cite{wengel}
or ground state calculations become mandatory. The latter program has
been performed in a tentative way in \cite{bokil} with a stochastic,
non-exact method for small system sizes ($L\le4$). The problem of
minimizing (\ref{vortex}) under the above mentioned constraints is a
convex cost flow problem that can be solved in a straightforward
manner with the algorithms presented in appendix \ref{app_convex}.
Work in this direction is in progress \cite{rieger_vortex}.

A further application of the minimum cost flow algorithms with convex
cost functions is {\bf traffic flow}, which became a major research topic in
{\it physics} quite recently \cite{traffic}. Network flow problems
naturally occur in any transportation system: what is the shortest
path between point A and point B in a road network (shortest path
problem), how many vessels does a steamship company need to have in
order to deliver perishable goods between several different
origin--destination pairs (maximum flow problem) or what is the flow
that satisfies the demands at a number of warehouses from the
available supplies at a number of plants and that minimizes its
shipping cost (typical transportation problem = minimum cost flow
problem).

All of the above problems are {\it linear} problems. Whenever system
congestion or queuing effects have to be taken into account in the
model describing a "real" network flow, the introduction of nonlinear
costs (since queuing delays vary nonlinearily with flows) are
mandatory. In road networks, as more vehicles use any road segment,
the road becomes increasingly congested and so the delay on that road
increases. For example, the delay on a particular road segment, as a
function of the flow $x$ on that road, might be $\alpha x/(u-x)$. In
this expression $u$ denotes a theoretical capacity of the road and
$\alpha$ is another constant: As the flow increases, so does the delay;
moreover, as the flow $x$ approaches the theoretical capacity of that
road segment, the delay on the link becomes arbitrarily large. In many
instances, as in this example, the delay function on each road segment
is a {\it convex} function of the road segment's flow, so finding the
flow plan that achieves the minimum overall delay, summed over all
road segments, is a convex cost network flow model.

It should have become clear at the end of this lecture that
frustrated, disordered systems and network flows are strongly related,
or even completely equivalent. The quenched disorder occurring in the
physical models we discussed so far find their counterpart in arc
capacities and costs in flow problems. Thus, many ``daily life''
networks, like transportation systems or urban traffic flows, share
many features with disordered or even glassy systems. For instance the
concept of {\it chaos} we encountered in spin glasses as well as in
the random solid-on-solid model should also be valid in traffic
networks: the slightest random (i.e.\ uncontrollable) change in the
capacities of the roads, as for instance after a heavy rain or
snowfall, or locally by several accidents, should completely change
the traffic flow pattern beyond a particular length scale. A
systematic study of these issues is certainly of great interest, not
only for the theoretical understanding of the intrinsically chaotic
nature of complex network flows but also for practical reasons.
\bigskip

\noindent
{\bf Acknowledgement:} I would like to express my special thanks to M.
J\"unger, U.\ Blasum, M.\ Diehl and N.\ Kawashima, from whom I learnt
about various aspects of the issues treated in this lecture and with
whom I enjoy(ed) an ongoing fruitful and lively collaboration. This work
has been supported by the Deutsche Forschungsgemeinschaft (DFG).
\vfill 
\eject

\appendix

\bc
\parbox{12cm}{{\LARGE\bf Concepts in network flows and basic algorithms}}
\ec

In this appendix we introduce the basic definitions in the theory of network
flows, which are needed in the main text. It represents a very
condensed version of some chapters of the excellent book {\it Network
flows} by R.\ K.\ Ahuja, T.\ L.\ Magnati and J.\ B.\
Orlin\cite{ahuja}.  The content of the subsequent chapters is
self-contained, so that it should be possible for the reader to
understand the basic ideas of the various algorithms.  In principle he
should even be able to devise a particular implementation of one or
the other code, although I recommend to consult existing
public-domain (!) software libraries (e.g.\ \cite{leda}) first.
\bigskip

\section{The maximum flow / minimum cut problem}
\label{app_maxflow}

\subsection{Basic definitions}
\label{app_def}

\noindent
A {\bf capacitated network} is a graph $G=(N,A)$, where $N$ is the set
of nodes and A the set of arcs, with {\it nonnegative} capacities
$u_{ij}$ (which can also be infinite) associated with each arc
$(i,j)\in A$. In our first example of the random bond Ising model $N$
is simply the set of lattice sites (plus two extra nodes, see fig.\
1), $A$ the bonds between interacting sites and $u_{ij}$ the
ferromagnetic interaction strengths. Note that $u_{ij}\ge0$ is
essential. Let $n=\#N$ be the number of nodes in $G$ and $m=\#A$ the
number of arcs.

\noindent
The {\bf arc adjacency list} is the set of arcs emanating from a node:
$A(i)=\{(i,j)|(i,j)\in A\}$.\\
\noindent
One distinguishes two special nodes of $N$:
the {\bf source} node $\s$ and the {\bf sink} node $\t$.

\noindent
A {\bf flow} in the network $G$ is a set of nonnegative numbers
$x_{ij}$ (or a map ${\bf x}:A\to R_+\cup\{0,\infty\}$) subject 
to a {\bf capacity constraint} for each arc
\be
0\le x_{ij}\le u_{ij}\qquad \forall(i,j)\in A\;.
\label{cap_constr}
\ee
and to a {\bf mass balance constraint} for each node
\be
\sum_{\{j|(i,j)\in A\}} x_{ij} - 
\sum_{\{j|(j,i)\in A\}} x_{ji} =
\left\{
\ba{rl}
-v & \quad{\rm for}\;i=\s\\
+v & \quad{\rm for}\;i=\t\\
0  & \quad{\rm else}
\ea
\right.
\label{flow}
\ee
This means that at each node everything that goes in has to go out, too,
with the only exception being the source and the sink. 
What actually flows from $\s$ to $\t$
is $v$, the value of the flow.

The {\bf maximum flow problem} for the capacitated network $G$ is
simply to find the flow ${\bf x}$ that has the maximum value $v$ under
the constraint (\ref{cap_constr}) and (\ref{flow}).

We make a few assumptions: {\bf 1)} the network is directed, which
means that for instance $(i,j)$ is an arc pointing from node $i$ to
node $j$ , {\bf 2)} whenever an arc $(i,j)$ belongs to a network, the
arc $(j,i)$ also belongs to it or is added with zero
capacity, {\bf 3)} all capacities are
nonnegative integers, {\bf 4)} the network does not contain a directed
path from node $\s$ to node $\t$ composed only of infinite capacity
arcs, {\bf 5)} the network does not contain parallel
arcs.\footnote{All of these assumptions can be fulfilled in the
  physical problems we consider by appropriate modifications. E.g.\ 
  number 3) can be fulfilled by rescaling the bond strengths $J_{ij}$
  with a factor and chopping off the decimal
  digits.}

\subsection{Residual Network and generic augmenting path algorithm}
\label{app_aug}

Now that we have defined the maximum flow problem, we have to
introduce some tools with which it can be solved. The most important
one is the notion of a {\it residual network}, which, as it is very
often in mathematics, is already half the solution. If we have found a
set of numbers $x$ that fulfill the mass balance constraints, we would
like to know whether this is already optimal, or on which arcs of the
network we can improve (or {\it augment} in the jargon of
combinatorial optimization) the flow.

\noindent
Given a flow ${\bf x}$, the {\bf residual capacity} $r_{ij}$ of any arc
$(i,j)\in A$ is the maximum additional flow that can be sent from
node $i$ to node $j$ using the arcs $(i,j)$ and $(j,i)$. The
residual capacity has two components: 1) $u_{ij}-x_{ij}$, the unused
capacity of arc $(i,j)$, 2) $x_{ji}$ the current flow on arc $(j,i)$,
which we can cancel to increase the flow from node $i$ to $j$.
\be
r_{ij}=u_{ij}-x_{ij}+x_{ji}
\label{red_cost}
\ee
The {\bf residual network} $G({\bf x})$  with respect to the flow ${\bf x}$
consists of the arcs with {\it positive} residual capacities.

\noindent
An {\bf augmenting path} is a directed path from the node $\s$ to the
node $\t$ in the residual network. The {\it capacity of an augmenting path}
is the minimum residual capacity of any arc in this path.

Obviously, whenever there is an augmenting path in the residual
network $G({\bf x})$ the flow ${\bf x}$ is not optimal. This motivates
the following generic augmenting path algorithm.
\bc
\parbox{11cm}{
\noindent
{\tt
{\bf algorithm} augmenting path\\
{\bf begin}\\
\spacea x:=0\\
\spacea {\bf while} 
         $G(x)$ contains a directed path from node $\s$ to $\t$
         {\bf do}\\
\spacea {\bf begin}\\
\spaceb identify an augmenting path $P$ from node $\s$ to node $t$\\
\spaceb $\delta={\rm min}\{r_{ij}|(i,j)\in P\}$\\
\spaceb augment $\delta$ units of flow along $P$ and update $G({\bf x})$\\
\spacea {\bf end}\\
{\bf end}
}}
\ec
Further below we will see that the flow is indeed maximal if there is
no augmenting path left. The main task in an actual implementation of
this algorithm would be the identification of the directed paths from
$\s$ to $\t$ in the residual network. Before we come to this point we
have to make the connection to the minimum cut problem that is
relevant for the physical problems discussed in the main text.

\subsection{Cuts, labeling algorithm and max-flow-min-cut theorem}
\label{app_mincut}

\noindent
A {\bf cut} is a partition of the node set $N$ into two subsets $S$
and $\overline{S}=N\backslash S$ denoted by $[S,\overline{S}]$.  We
refer to a cut as a {\bf $\s$-$\t$-cut} if $\s\in S$ and
$\t\in\overline{S}$.

\noindent
The {\bf forward} arcs of the cut $[S,\overline{S}]$ are those
arcs $(i,j)\in A$ with $i\in S$ and $j\in\overline{S}$, the 
{\bf backward} arcs those with $j\in S$ and $i\in\overline{S}$.
The set of all forward arcs of $[S,\overline{S}]$ is denoted
$(S,\overline{S})$.

\noindent
The {\bf capacity} of an $\s$-$\t$-cut is defined to be
$u[S,\overline{S}]=\sum_{(i,j)\in(S,\overline{S})} u_{ij}$.
Note that the sum is only over forward arcs of the cut.

\noindent
The {\bf minimum cut} is a $\s$-$\t$-cut whose capacity $u$ is
minimal among all $\s$-$\t$-cuts.

Let ${\bf x}$ be a flow, $v$ its value and $[S,\overline{S}]$ an
$\s$-$\t$-cut. Then, by adding the mass balances for all nodes in $S$
we have
\be
v=\sum_{i\in S}\biggl\{
\sum_{\{j|(i,j)\in A(i)\}} x_{ij} - 
\sum_{\{j|(j,i)\in A(i)\}} x_{ji} \biggr\}
=\sum_{(i,j)\in(S,\overline{S})} x_{ij}
-\sum_{(i,j)\in(\overline{S},S)} x_{ji}
\;.
\ee
Since $x_{ij}\le u_{ij}$ and $x_{ji}\ge0$ the following inequality holds
\be
v\le\sum_{(i,j)\in(S,\overline{S})} u_{ij}=u[S,\overline{S}]
\ee
Thus the value of any flow ${\bf x}$ is less or equal to the capacity
of any cut in the network. If we discover a flow ${\bf x}$ whose value
equals to the capacity of some cut $[S,\overline{S}]$, then ${\bf x}$
is a maximum flow and the cut is a minimum cut.  The following
implementation of the augmenting path algorithm constructs a flow
whose value is equal to the capacity of a $\s$-$\t$-cut it defines
simultaneously. Thus it will solve the maximum flow problem (and, of
course, the minimum cut problem).

As we have mentioned, our task is to find augmenting paths in the
residual network. The following {\bf labeling algorithm} uses a
search technique to identify a directed path in $G({\bf x})$ from the
source to the sink. The algorithm fans out from the source node to
find all nodes that are reachable from the source along a directed
path in the residual network. At any step the algorithm has
partitioned the nodes in the network into two groups: {\it labeled}
and {\it unlabeled}. The former are those that the algorithm was able
to reach by a directed path from the source, the latter are those that
have not been reached yet. If the sink becomes labeled the algorithm
sends flow along a path (identified by a predecessor list) from $\s$
to $\t$. If all labeled nodes have been scanned and it was not
possible to reach the sink, the algorithm terminates.
\vfill
\eject

\bc
\parbox{\hsize}{
\noindent
{\tt
{\bf algorithm} labeling\\
{\bf begin}\\
\spacea label node $\t$\\
\spacea {\bf while} node $\t$ is labeled {\bf do}\\
\spacea {\bf begin}\\
\spaceb unlabel all nodes\\
\spaceb set $pred(j)=0$ for each $j\in N$\\
\spaceb label node $\s$ and set $list:=\{\s\}$\\
\spaceb {\bf while} $list\ne\emptyset$ and node $\t$ is unlabeled {\bf do}\\
\spaceb {\bf begin}\\
\spacec remove a node $i$ from $list$\\
\spacec {\bf for} each arc $(i,j)\in A(i)$ in the residual network {\bf do}\\
\spaced {\bf if} $r_{ij}>0$ and node $j$ is unlabeled {\bf then}\\
\spacee set $pred(j)=i$\\
\spacee label node $j$\\
\spacee add node $j$ to $list$\\
\spaceb {\bf end}\\
\spaceb {\bf if} node $\t$ is labeled {\bf then} $augment$\\
\spacea {\bf end}\\
{\bf end}
}}
\ec

\bc
\parbox{\hsize}{
\noindent
{\tt
{\bf procedure} $augment$\\
{\bf begin}\\
\spacea Use the predecessor labels to trace back from the sink to\\
\spacea the source to obtain an augmenting path $P$ from $\s$ to $\t$\\
\spacea $\delta={\rm min}\,\{r_{ij}|(i,j)\in P\}$\\
\spacea augment $\delta$ units of flow along $P$, update residual capacities\\
{\bf end}
}}
\ec
Note that in each iteration the algorithm either performs an
augmentation or terminates because it cannot label the sink. In the
latter case the current flow is a maximum flow. To see this, suppose
that at this stage $S$ is the set of labeled nodes and
$\overline{S}=N\backslash S$ is the set of unlabeled nodes. Clearly
$\s\in S$ and $\t\in\overline{S}$. Since the algorithm cannot label
any node in $\overline{S}$ from any node in $S$, $r_{ij}=0$ for each
$(i,j)\in(S,\overline{S})$, which implies with (\ref{red_cost})
$x_{ij}=u_{ij}+x_{ji}$. Thus $x_{ij}=u_{ij}$ (since $0\le x_{ij}\le
u_{ij}$) for all $(i,j)\in(S,\overline{S})$ and $x_{ij}=0$ for all
$(i,j)\in(\overline{S},S)$. Hence
\be
v=\sum_{(i,j)\in(S,\overline{S})} x_{ij}
-\sum_{(i,j)\in(\overline{S},S)} x_{ij}
=\sum_{(i,j)\in(S,\overline{S})} x_{ij}=u[S,\overline{S}]\;.
\ee
This means that the flow ${\bf x}$ equals the capacity of the cut
$[S,\overline{S}]$, and therefore ${\bf x}$ is a {\bf maximum flow} and
$[S,\overline{S}]$ is a {\bf minimum cut}.

From these observation let us note the following conclusions:

\noindent 
{\bf Max-flow-min-cut theorem}: The maximum value of the
flow from a source node $\s$ to a sink node $\t$ in a capacitated
network equals the minimum capacity among all $\s$-$\t$-cuts.

\noindent
{\bf Augmenting path theorem}: A flow ${\bf x^*}$ is a maximum flow if and
only if the residual network $G({\bf x^*})$ contains no augmenting
path.

\noindent
{\bf Integrality theorem}: If all arc capacities are integer, the
maximum flow problem has an integer maximum flow.

Let $n$ be the number of nodes, $m$ the number of arcs and $U={\rm
max}\{ u_{ij} \}$. Since any arc is at most examined once and the cut
capacity is at most $nU$ the complexity of this algorithm is ${\cal
O}(nmU)$ (note that the flow increases at least by $1$ in each
augmentation). Because of the appearance of the number $U$ it is a
pseudo-polynomial algorithm. The so called preflow-push algorithms we
discuss now are much more efficient, in particular they avoid the
delay caused notoriously by some bottleneck situations.

\subsection{Preflow-push algorithm}
\label{app_push}

The inherent drawback of the augmenting path algorithms is the
computationally expensive operation of sending flow along a path,
which requires ${\cal O}(n)$ time in the worst case. The preflow-push
algorithms push flow on individual arcs instead of augmenting
paths. They do not satisfy the mass balance constraints at
intermediate stages. This is a very general concept in combinatorial
optimization: algorithms either can operate within the space of
admissible solutions and try to find optimality during iteration, or
they can fulfill some sort of optimality criterion all the time and
strive for feasibility. Augmenting path algorithms are of the first
kind, preflow-push algorithms of the second. The basic idea is to
flood the network from the source pushing as much flow as the arc
capacities allow into the network towards the sink and then reduce it
successively until the mass balance constraints are fulfilled.

A {\bf preflow} is a function ${\bf x}:A\to R$ that satisfies the flow
bound constraint $x_{ij}\le u_{ij}$ and the following relaxation for
the {\bf excess} $e(i)$ of each node $i$:
\be
e(i):=
\sum_{\{j|(j,i)\in A\}} x_{ji} - 
\sum_{\{j|(i,j)\in A\}} x_{ij} \ge0
\qquad\forall i\in N\backslash\{\s,\t\}\;.
\ee
It is $e(\t)\ge0$ and only $e(\s)<0$. One denotes a node $i$ to be
{\bf active} if its excess is strictly positive $e(i)>0$.

As mentioned, preflow-push algorithms strive to achieve
feasibility. Active nodes indicate that the solution is
infeasible. Thus the basic operation of the algorithm is to select an
active node and try to remove its excess by pushing flow to its
neighbors. Since ultimately we want to send flow to the sink, we push
flow to the nodes that are {\it closer} to the sink. This necessitates
the use of distance labels:

\noindent
We say that a {\bf distance function} $d:N\to Z_+\cup\{0\}$ is {\bf
valid} with respect to a flow ${\bf x}$, if it satisfies\\
\spacea a) $d(\t)=0$ and\\
\spacea b) $d(i)\le d(j)+1$ for every arc $(i,j)$ in the residual network $G({\bf x})$.\\

\noindent
If $d(\cdot)$ is valid then it has also the following properties
(where $n$ is the number of nodes):\\
\parbox{15cm}{
\spacea 1) $d(i)\le$ length of the shortest directed path from 
           node $i$ to $\t$ in $G({\bf x})$\\
\spacea 2) $d(\s)\ge n\Rightarrow$ $G({\bf x})$ contains no 
            directed path from $\s$ to $\t$.}
Furthermore we say that $d(\cdot)$ is {\bf exact} if in 
1) the equality holds.\\
Finally an arc $(i,j)$ is {\bf admissible} if $d(i)=d(j)+1$.

In the preflow-push algorithm we push flow on these admissible
arcs. If the active node that we are currently considering has no
admissible arcs, we increase its distance label so that we create at
least one admissible arc.
\bc
\parbox{10.4cm}{
\noindent
{\tt
{\bf algorithm} preflow-push\\
{\bf begin}\\
\spacea {\it preprocess}\\
\spacea {\bf while} the network contains an active node {\bf do}\\
\spacea {\bf begin}\\
\spaceb select an active node $i$\\
\spaceb {\it push/relabel(i)}\\
\spacea {\bf end}\\
{\bf end}
}}
\ec

\bc
\parbox{10.4cm}{
\noindent
{\tt
{\bf procedure} {\it preprocess}\\
{\bf begin}\\
\spacea ${\bf x}:=0$\\
\spacea compute the exact distance labels $d(i)$\hfill{\bf(1)}\\
\spacea $x_{\s j}=u_{\s j}$ for each arc $(\s,j)\in A$\\
\spacea $d(\s)=n$\\
{\bf end}
}}
\ec

\bc
\parbox{10.4cm}{
\noindent
{\tt
{\bf procedure} {\it push/relabel(i)}\\
{\bf begin}\\
\spacea {\bf if} the network contains an admissible arc $(i,j)$ {\bf then}\\
\spaceb push $\delta={\rm min}\,\{e(i),r_{ij}\}$ units of flow from node $i$ to $j$\\
\spacea {\bf else}\\
\spaceb replace $d(i)$ by ${\rm min}\,\{d(j)+1|(i,j)\in A\;{\rm and}\;r_{ij}>0\}$\\
{\bf end}
}}
\ec

Ad {\bf (1)}: To compute the exact distance labels we have to
calculate the shortest distances from node $\t$ to every other node,
which we learn how to do in the next section.

The algorithm terminates when the excess resides at the source or at
the sink, implying that the current preflow is a {\it flow}. Since
$d(\s)=n$ after peprocessing, and $d(i)$ is never decreased in
push/relabel(i) for any $i$, the residual network contains no path
from $\s$ to $\t$, which means according to 2) above that there is no
augmenting path. Thus the flow is maximal.

As in the context of the max-flow-min-cut theorem of the last section
it might also here be instructive to visualize the generic
preflow-push algorithm in terms of a network of (now flexible) water
pipes representing the arcs with joints being the nodes. The distance
function, which is so essential in this algorithm, measures how far
nodes are above the ground. In this network we wish to send water from
the source to the sink. In addition we visualize flow in admissible
arcs as water flowing downhill. Initially, we move the source node
upward, and water flows to its neighbors. In general, water flows
downwards to the sink; however, occasionally flow becomes trapped
locally at a node that has no downhill neighbors. At this point we
move the node upward, and again water flows downhill to the
sink. Eventually, no more flow can reach the sink. As we continue to
move nodes upward, the remaining excess flow eventually flows back
towards the source. The algorithm terminates when all the water
flows either into the sink or flows back to the source.

The complexity of this algorithm turns out to be ${\cal O}(n^2m)$, the
so called FIFO preflow-push algorithm, which we do not discuss here,
has a complexity of ${\cal O}(n^3)$.

\section{Shortest path problems}
\label{app_short}

\subsection{Dijkstra's algorithm}
\label{app_dijk}

Given a network $G=(N,A)$ and for each arc $(i,j)\in A$ a non-negative
arc-length $c_{ij}$. In the above problem, where we had to find the
exact distance labels in the preflow-push algorithm it is simply
$c_{ij}=1$ for all arcs in the residual network.

The task is to find the shortest paths from one particular node $\s$
to all other nodes in the network. {\it Dijkstra's algorithm} is a
typical label-setting algorithm to solve this problem (with complexity
${\cal O}(n^2)$. It provides distance labels $d(i)$ with each node. Each
of these is either temporarily (defining a set $S$) or permanently
(defining a set $\overline{S}=N\backslash S$) labeled during the
iteration, and as soon as a node is permanently labeled, $d(i)$ is
the shortest distance. The path itself is reconstructed via
predecessor indices.

First note that $d(j)=d(i)+c_{ij}$ for each arc $(i,j)$ in a shortest
path from node $\s$ to node $i$, and that $d(j)\ge d(i)+c_{ij}$
otherwise. By fanning out from node $\s$ the algorithm uses this
criterion to find successively the shortest paths.
\bc
\parbox{10cm}{
\noindent
{\tt
{\bf algorithm} Dijkstra\\
{\bf begin}\\
\spacea $S:=\emptyset$, $\overline{S}=N$\\
\spacea $d(i):=\infty$ for each node $i\in N$\\
\spacea $d(\s):=0$ and $pred(\s):=0$\\
\spacea {\bf while} $|S|<n$ {\bf do}\\
\spacea {\bf begin}\\
\spaceb let $i\in\overline{S}$ be a node for which $d(i)={\rm min}\,
            \{d(j)|j\in\overline{S}\}$\\
\spaceb $S:=S\cup\{i\}$, $\overline{S}:=\overline{S}\backslash\{i\}$\\
\spaceb {\bf for} each $(i,j)\in A(i)$ {\bf do}\\
\spacec {\bf if} $d(j)>d(i)+c_{ij}$ {\bf then}\\
\spaced $d(j):=d(i)+c_{ij}$ and $pred(j):=i$\\
\spacea {\bf end}\\
{\bf end}
}}
\ec
The fact that we always add a node $i\in\overline{S}$ with {\it
  minimal} distance label $d(i)$ ensures that $d(i)$ is indeed a
shortest distance (there might be other shortest paths, but none with
a strictly shorter distance). There are special implementations of
this algorithm that have a much better running time than 
${\cal O}(n^2)$.

\subsection{Label correcting algorithm}
\label{app_labelcorr}

As we said, Dijkstra's algorithm is a label-setting algorithm.  The
above mentioned criterion 
\bc
$d(i)$ shortest path distances $\;\Leftrightarrow\;$
$d(j)\le d(i)+c_{ij}$ $\forall\,(i,j)\in A$
\ec
gives also rise to a so called {\it label-correcting} algorithm.\\
Let us define reduced arc length (or {\bf reduced costs}) via
\be
c_{ij}^d:=c_{ij}+d(i)-d(j)\;.
\ee
As long as one reduced arc lengths is negative, the distance
labels $d(i)$ are not shortest path distances:
\be
d(\cdot)\;{\rm shortest\;path\;distances}\quad
\Leftrightarrow 
\quad c_{ij}^d\ge0\qquad\forall(i,j)\in A
\label{dist_crit}
\ee
For later reference we also note the following observation. For any
directed cycle $W$ one has
\be
\sum_{(i,j)\in W} c_{ij}^d=\sum_{(i,j)\in W} c_{ij}
\ee
The criterion (\ref{dist_crit}) suggests the following algorithm
for the shortest path problem:

\bc
\parbox{11cm}{
\noindent
{\tt
{\bf algorithm} label-correcting\\
{\bf begin}\\
\spacea $d(\s):=0$ and $pred(\s):=0$\\
\spacea $d(j):=\infty$ for each node $j\in N\backslash\{\s\}$\\
\spacea {\bf while} some arc $(i,j)$ satisfies 
        $d(j)>d(i)+c_{ij}$ ($c_{ij}^d<0$) {\bf do}\\
\spacea {\bf begin}\\
\spaceb $d(j):=d(i)+c_{ij}$ ($\Rightarrow c_{ij}^d=0$)\\
\spaceb $pred(j)=i$\\
\spacea {\bf end}\\
{\bf end}
}
}
\ec
The generic implementation of this algorithm has a running time ${\cal
  O}({\rm min}\{n^2mC,$ $m2^n\})$ with $C={\rm max}\,|c_{ij}|$, which is
pseudo-polynomial.  A FIFO implementation has complexity ${\cal O}(nm)$.

This algorithm also works for the cases in which some costs $c_{ij}$
are negative, provided there are {\it no negative cycles}, i.e.\
closed directed paths $W$ with $\sum_{(i,j)\in W} c_{ij}<0$. In that
case the instruction $d(j):=d(i)+c_{ij}$ would decrease some distance
labels {\it ad (negative) infinitum}.

If there {\it are} negative cycles, one can detect them with an appropriate
modification of the above code: One can terminate if $d(k)<-nC$ for
some node $k$ (again $C={\rm max}\,|c_{ij}|$) and obtain these
negative cycles by tracing them through the predecessor indices
starting at node $k$. This will be useful in the next section.
\vfill

\section{Minimum cost flow problems}
\label{app_mincost}

\subsection{Definition}
\label{app_def2}

The next flow problem we discuss combines features of the maximum-flow
and the shortest paths problem. The algorithm that solves it therefore
also makes use of the ideas we presented so far. Let $C=(N,A)$ be a
directed network with a {\it cost} $c_{ij}$ and a {\it capacity}
$u_{ij}$ associated with every arc $(i,j)\in A$. Moreover we associate
with each node $i\in N$ a number $b(i)$ which indicates its {\it
supply} or {\it demand} depending on whether $b(i)>0$ or $b(i)<0$. The
{\it minimum cost flow problem} is
\be
{\rm Minimize}\quad z({\bf x})=\sum_{(i,j)\in A} c_{ij} x_{ij}
\label{mc_min}
\ee
subject to the mass balance constraints
\be
\sum_{\{j|(i,j)\in A\}} x_{ij} - 
\sum_{\{j|(j,i)\in A\}} x_{ji} = b(i) \qquad\forall i\in N
\label{mc_mass}
\ee
and the capacity constraints
\be
0\le x_{ij}\le u_{ij}\qquad\forall (i,j)\in A
\label{mc_cap}
\ee
Again we make a few assumptions: {\bf 1)} All data (cost,
supply/demand, capacity) are integral\footnote{Here the same remark
  holds as for the maximum flow problem, previous footnote.}, {\bf 2)}
the network is directed, {\bf 3)} $\sum_i b(i)$ and the minimum cost
flow problem has a feasible solution (that means, one can find a flow
$x_{ij}$ that fulfills the mass balance and capacity
constraints\footnote{In the physical models we discuss it is $b(i)=0$
  anyway, implying ${\bf x}=0$ as a feasible solution.}, {\bf 4)} it
exists an uncapacitated directed path between every pair of nodes,
{\bf 5)} all arc costs are non-negative (otherwise one could
appropriately define a revered arc).

Again the {\bf residual network} $G({\bf x})$ corresponding to a flow
${\bf x}$ will play an essential role. It is defined in the same way as
in the maximum flow problem, in addition the costs for the
backwards arcs are reversed: a flow $x_{ij}$ on arc $(i,j)\in A$ with
capacity $u_{ij}$ and cost $c_{ij}$ will give rise to the arcs $(i,j)$
and $(j,i)$ with residual capacities $r_{ij}=u_{ij}-x_{ij}$ and
$r_{ji}=x_{ij}$, respectively and costs $c_{ij}$ and $-c_{ij}$
respectively.

\subsection{Negative cycle canceling algorithm}
\label{app_negcyc}

First we formulate a very important intuitive optimality criterion, the
{\bf negative cycle optimality criterion}: A feasible solution ${\bf
x^*}$ is an {\it optimal} solution of the minimum cost flow problem,
{\it if and only if} the residual network $G({\bf x^*})$ contains {\it
no negative cost cycle}.

The proof is easy: Suppose the flow ${\bf x}$ is feasible and $G({\bf
x})$ contains a negative cycle. The a flow augmentation along this
cycle improves the function value $z({\bf x})$, thus ${\bf x}$ is not
optimal. Now suppose that ${\bf x^*}$ is feasible and $G({\bf x^*})$
contains no negative cycles and let ${\bf x^0}\ne{\bf x^*}$ be an
optimal solution. Now decompose ${\bf x^0}-{\bf x^*}$ into augmenting
cycles, the sum of the costs along these cycles is ${\bf c\cdot
x^0}-{\bf c\cdot x^*}$. Since $G({\bf x^*})$ contains no negative
cycles ${\bf c\cdot x^0}-{\bf c\cdot x^*}\ge0$, and therfore ${\bf
c\cdot x^0}={\bf c\cdot x^*}$ because optimality of ${\bf x^*}$
implies ${\bf c\cdot x^0}\le{\bf c\cdot x^*}$. Thus ${\bf x^0}$ is
also optimal.

The following algorithm iterativly cancels negative cycles until the
optimal solution is reached.
\bc
\parbox{11cm}{
\noindent
{\tt
{\bf algorithm} cycle canceling\\
{\bf begin}\\
\spacea establish a feasible flow {\bf x}\hfill{\bf(1)}\\
\spacea {\bf while} $G({\bf x})$ contains a negative cycle {\bf do}\\
\spacea {\bf begin}\\
\spaceb use some algorithm to identify a negative cycle $W$\hfill{\bf(2)}\\
\spaceb $\delta:={\rm min}\,\{r_{ij}|(i,j)\in W\}$\\
\spaceb augment $\delta$ units of flow in the cycle and update $G({\bf x})$\\
\spacea {\bf end}\\
{\bf end}
}}
\ec

Ad {\bf (1)}: Although, as we mentioned, in the physical problems we
discuss a feasible solution is obvious in most cases (e.g.\ ${\bf
x}=0$) we note that in principle one has to solve a maximum flow
problem here: One introduces two extra-nodes $\s$ and $\t$ (source and
sink, of course) and\\ 
$\forall i:b(i)>0$ add a source arc $(\s,i)$ 
with capacity $u_{\s i}=b(i)$\\ 
$\forall i:b(i)<0$ add a sink arc
$(i,\t)$ with capacity $u_{i\t}=-b(i)$.\\ 
If the maximum flow from $\s$ to $\t$ saturates all source arcs
(remember $\sum_i b(i)=0$) the minimum cost flow problem is feasible
and the maximum flow ${\bf x}$ is a feasible flow.

Ad {\bf (2)}: For negative cycle detection in the residual network
$G(x)$ one can use the label-correcting algorithm for the shortest
path problem presented in the last section.

The running time of this algorithm is ${\cal O}(mCU)$, where $C={\rm
  max}\,|c_{ij}|$ and $U={\rm max}\,u_{ij}$, which means that
it is pseudopolynomial. In the next section we present an alternative
and more efficient way to solve the minimum cost flow problem.

\subsection{Reduced cost optimality}
\label{app_redcost}

Remember that when we considered the shortest path problem we
introduced the reduced costs and obtained the shortest path optimality
condition $c_{ij}^d=c_{ij}+d(i)-d(j)\ge0$.  This means
\begin{itemize}
\item 
$c_{ij}^d$ is an optimal ``reduced cost'' for arc $(i,j)$ in the
sense that it measures the cost of this arc relative to the shortest
path distances.
\item
With respect to the optimal distances, every arc has a nonnegative
cost.
\item
Shortest paths use only zero reduced cost arcs.
\item
Once we know the shortest distances, the shortest path problem is easy
to solve: Simply find a path from node $\s$ to every other node using
only zero reduced cost arcs.
\end{itemize}
The natural question arises, whether there is a similar set of
conditions for more general min-cost flow problems. The answer is yes
as we show in the following.

\noindent
For the network defined in the last section associate a real number
$\pi(i)$, unrestricted in sign with each node $i$, $\pi(i)$ is the
{\bf potential} of node $i$.

\noindent
We define the {\bf reduced cost} of arc $(i,j)$ of a set of node
potentials $\pi(i)$
\be
c_{ij}^{\pi}:=c_{ij}-\pi(i)+\pi(j)\;.
\ee
The reduced costs in the {\it residual network} are defined in the
same way as the costs, but with $c_{ij}^\pi$ instead of $c_{ij}$.

\noindent
We have\\ 1) For any directed path $P$ from $k$ to $l$:
$\sum_{(i,j)\in P}c_{ij}^\pi=\sum_{(i,j)\in P}c_{ij}+d(k)-d(l)$.\\ 2)
For any directed cycle $W$: $\sum_{(i,j)\in
W}c_{ij}^\pi=\sum_{(i,j)\in W}c_{ij}$\\ This means that negative
cycles with respect to $c_{ij}$ are also negative cycles with respect
to $c_{ij}^\pi$.

Now we can formulate the reduced cost optimality condition:
\bc
A feasible solution ${\bf x^*}$ is an optimal solution of the min-cost
flow problem\\
$\Leftrightarrow$\\
\parbox{12cm}{$\exists\,\pi$, a set of node potentials that satisfy the reduced 
cost optimality condition}\\
$c_{ij}^\pi\ge0\qquad\forall(i,j)$ arc in $G({\bf x^*})$.
\ec
For the implication ``$\Leftarrow$'' suppose that
$c_{ij}^\pi\ge0\,\forall(i,j)$. One immediately realizes that
$G({\bf x^*})$ contains no negative cycles since for each cycle $W$
one has $\sum_{(i,j)\in W}c_{ij}=\sum_{(i,j)\in W}c_{ij}^\pi\ge0$.
For the other direction ``$\Rightarrow$'' suppose that $G({\bf x^*})$
contains no negative cycles. Denote with $d(\cdot)$ the shortest path
distances from node 1 to all other nodes. Hence $d(j)\le d(i)+c_{ij}$
$\forall(i,j)\in G({\bf x^*})$. Now define $\pi=-d$ then
$c_{ij}^\pi=c_{ij}+d(i)-d(j)\ge0$. Note how closely connected the
shortest path problem is to the min-cost flow problem.

There is an intuitive economic interpretation of the reduced cost
optimality condition. Suppose we interprete $c_{ij}$ as the cost of
transporting one unit of a commodity from node $i$ to node $j$ through
arc $(i,j)$ and $\mu(i)=-\pi(i)$ as the cost of {\it obtaining} it at
$i$. Then $c_{ij}+\mu(i)$ is the cost of commdity at node $j$, if we
obtain it at node $i$ and transport it to node $j$ via arc
$(i,j)$. The inequality $c_{ij}^\pi\ge0\,\Leftrightarrow\mu(j)\le
c_{ij}+\mu(i)$ says that the cost of commodity at node $j$ is no more
than obtaining it at $i$ and sending it via $(i,j)$ --- it could be
smaller via other paths.
\vfill
\eject

\subsection{Successive shortest path algorithm}
\label{app_succ}

With the concept of reduced costs we can now introduce the successive
shortest path algorithm for solving the min-cost flow problem. The
cycle canceling algorithm maintains feasibility of the solution at
every step and attempts to achieve optimality. In contrast, the
successive shortest path algorithm maintains optimality of the
solution ($c_{ij}^\pi\ge0$) at every step and strives to attain
feasibility (with respect to the mass balance constraints).

\noindent
A {\bf pseudoflow} ${\bf x}:A\to R^+$ satisfies the capacity and
non-negaivity constraints, but not necessarily the mass balance
constraints.

\noindent
The {\bf imbalance} of node $i$ is defined as
\be
e(i):=b(i)+\sum_{\{j|(ji)\in A\}} x_{ji}
-\sum_{\{j|(ji)\in A\}} x_{ij}.
\ee
If $e(i)>0$ then we call $e(i)$ the {\bf excess} of node $i$, If
$e(i)<0$ then we call it the {\bf deficit}. $E=\{i|e(i)>0\}$ and
$E=\{i|e(i)<0\}$ are the sets of excess and deficit nodes,
respectively. Note that because of $\sum_{i\in N} e(i)=\sum_{i\in N}
b(i)=0$ we have $\sum_{i\in E} e(i)=-\sum_{i\in D} e(i)$.

Let the pseudoflow {\bf x} satisfy the reduced cost optimality
condition with respect to the node potential $\pi$ and $d(\cdot)$ the
shortest path distances from some node $\s$ to all the other nodes in
the residual network $G({\bf x})$ with $c_{ij}^\pi$ as arc
lengths. Therefore we have:
\medskip

\noindent
{\bf Lemma 1}:\\
\spacea
{\bf a)} For the potential $\pi'=\pi-d$ we have $c_{ij}^{\pi'}\ge0$, too.\\
\spacea
{\bf b)} $c_{ij}^{\pi'}=0$ for all arcs $(i,j)$ on shortest paths.\\
To see a) note that 
$d(j)\le d(i)+c_{ij}^\pi$, thus 
$c_{ij}^{\pi'}=c_{ij}+(\pi(i)-d(i))-(\pi(j)-d(i))=c_{ij}^\pi+d(j)-d(i)\ge0$.
For b) replace only the inequality by an equality.
\medskip

The following lemma is the basis of the subsequent algorithm: Make the
same assumption as in Lemma 1. Now send flow along a shortest path
from some node $\s$ to some other node $k$ to obtain a new pseudoflow
${\bf x'}$.
\medskip

\noindent
{\bf Lemma 2:}\\
\spacea ${\bf x'}$ also satisfies the reduced
cost optimality conditions!
\medskip

\noindent
For the proof take $\pi$ and $\pi'$ as
in Lemma 1 and let $P$ be the shortest path from node $\s$ to node
$k$. Because of part b) of Lemma 1 it is $\forall\,(i,j)\in P:\;
c_{ij}^{\pi'}=0$. Therefore
$c_{ji}^{\pi'}=-c_{ij}^{\pi'}=0$. Thus a flow augmentation on
$(i,j)\in P$ might add $(j,i)$ to the residual network, but
$c_{ji}^{\pi'}=0$, which means that still the reduced cost optimality
condition $c_{ji}^{\pi'}\ge0$ is fulfilled.

The strategy for an algorithm is now clear. By starting with a
feasible solution that fulfills the reduced cost optimality condition
one can iteratively send flow along the shortest paths from excess
nodes to deficit nodes to finally fulfill also the mass balance
constraints.
\vfill

\bc
\parbox{\hsize}{
\noindent
{\tt
{\bf algorithm} successive shortest paths\\
{\bf begin}\\
\spacea ${\bf x}:=0$ ($G({\bf x})=G$) and $\pi:=0$ ($c_{ij}=c_{ij}^\pi\ge0$)\\
\spacea $e(i):=b(i)$ $\forall i\in N$\\
\spacea $E:=\{i|e(i)>0\}$, $D:=\{i|e(i)<0\}$.\\
\spacea {\bf while} $E\ne\emptyset$ {\bf do}\\
\spacea {\bf begin}\\
\spaceb select a node $k\in E$ and a node $l\in D$\\
\spaceb determine shortest path distance $d(j)$ from some node $\s$ to\\
\spacec all other nodes in $G({\bf x})$ w.r.\ to the reduced costs $c_{ij}^\pi$\\
\spaceb let $P$ denote a shortest path from node $k$ to node $l$\\
\spaceb update $\pi:=\pi-d$\\
\spaceb $\delta:={\rm min}\,\{e(k),-e(l),{\rm min}\,\{r_{ij}|(i,j)\in P\}\,\}$\\
\spaceb augment $\delta$ units of flow along path $P$\\
\spaceb update ${\bf x}$, $G({\bf x})$, $E$, $D$ and the reduced costs\\
\spacea {\bf end}\\
{\bf end}
}}
\ec
Note that in each iteration one excess is decreased by increasing flow
ny at least one unit. Denoting with $U$ the upper bound on the largest
supply of any node one needs at most $nU$ iterations, in each of which
one has to solve a shortest path problem with non-negative arc lengths
(so Dijkstra's algorithm is appropriate). This means that the above
algorithm is polynomial if we know how $U$ scales with $m$ or $n$.

\subsection{Convex cost flows}
\label{app_convex}

The cycle annealing algorithm as well as the successive shortest path
algorithm solve the minimum cost flow problem for a {\it linear} cost
function $\sum_{(i,j)\in E} c_{ij}\cdot x_{ij}$, where $c_{ij}$
represents the cost for sending one unit of flow along along the arc
$(i,j)$. This problem can be generalized to the following situation:
\be
{\rm Minimize}\quad z({\bf x})=\sum_{(i,j)\in A} h_{ij}(x_{ij})
\ee
subject to the mass balance constraints (\ref{mc_mass}) and the
capacity constraint (\ref{mc_cap}). In addition we demand the
flow variables $x_{ij}$ to be integer. The functions $h_{ij}(x_{ij})$ can
be any non-linear function, which has, however, to be {\it convex},
i.e.\
\be
\forall x,y\;,{\rm and}\;\theta\in[0,1]\quad 
h_{ij}(\theta x+(1-\theta)y)
\le \theta h_{ij}(x) + (1-\theta) h_{ij}(y)
\ee
For this reason it is called {\it convex cost flow problem}.
Without loss of generality we can assume that $h_{ij}(0)=0$.
Here the cost (for {\it one} unit) depends on the actual
flow (since $h_{ij}(x_{ij})$ is a nonlinear function of the flow
variable $x_{ij}$):
\be
\ba{lcl}
c_{ij}(x_{ij}) & := & h_{ij}(x_{ij}+1) - h_{ij}(x_{ij})\\
c_{ji}(x_{ij}) & := & h_{ij}(x_{ij}-1) - h_{ij}(x_{ij})
\label{costs}
\ea
\ee
Now $c_{ij}$ and $c_{ji}$ are the costs for increasing and decreasing,
respectively, the the flow variable $x_{ij}$ by one.

After introducing these quantities it becomes straightforward to solve
this problem with slight modifications of the algorithms we have
already at hand. The first is again a negative cycle canceling
algorithm:
\bc
\parbox{11cm}{
\noindent
{\tt
{\bf algorithm} cycle canceling (convex costs)\\
{\bf begin}\\
\spacea establish a feasible flow {\bf x}\\
\spacea calculate the costs $c(x)$ as in eq.\ (\ref{costs})\\
\spacea {\bf while} $[G({\bf x}),c(x)]$ contains a negative cycle {\bf
  do}\\
\spacea {\bf begin}\\
\spaceb use some algorithm to identify a negative cycle $W$\\
\spaceb augment {\it one} unit of flow in the cycle\\
\spaceb update $G({\bf x})$ and $c(x)$\\
\spacea {\bf end}\\
{\bf end}
}}
\ec
Note that since $h_{ij}(x)$ is convex the cost for augmenting $x$ by
{\it more} than one unit {\it increases} the costs. This ensures that
if we do not find any negative cycles, the flow is indeed optimal.

This algorithm is, unfortunately non-polynomial in time, although it
performs reasonably well on average. The successive shortest path
algorithm discussed in the last section can also be applied in the
present context with a significant gain in efficiency.  For this
algorithm it was essential that the reduced costs $c_{ij}^\pi$ with
respect to some node potential $\pi$ maintains the reduced cost
optimality condition $c_{ij}^\pi\ge0$ upon flow augmentation along
shortest paths. Now the question is, whether this still holds if with
the change of the flow (caused by the augmentation) also the costs
change. To show this we prove the folowing

\noindent
{\bf Lemma:} Let $\s$ be an excess node, $d(\cdot)$ shortest path
distances w.r.\ to the reduced costs $c_{ij}^\pi$ from node $\s$ to
all other nodes, $\pi'=\pi+d$, $\t$ a deficit node, $W$ a shortest
path from $\s$ to $\t$, and ${\bf x}^{\rm new}$ the flow that one
obtains by augmenting ${\bf x}$ along $W$ by {\it one} unit.
Then:
\bc
$c_{ij}^{\pi'}\ge0$ also for the {\it modified arc costs} along $W$.
\ec
For the proof let $w_{ij}\in\{+1,-1\}$ for an arc $(i,j)\in W$ with
$x_{ij}^{\rm new}=x_{ij}+w_{ij}$. Then the modified costs on this arc
are\\
$$
c_{ij}^*=h_{ij}(x_{ij}+2w_{ij})-h_{ij}(x_{ij}+w_{ij})\ge
h_{ij}(x_{ij}+w_{ij})-h_{ij}(x_{ij})=c_{ij}\;.
$$
because of the convexity of $h_{ij}(x)$.  
From this follows for the modified reduced costs
$c_{ij}^{\pi'*}=c_{ij}^{\pi'}+\pi'(i)-\pi'(j)\ge c_{ij}+\pi'(i)-\pi'(j)$,
which proves the lemma.

Thus we have the successive shortest path algorithm for the convex
costs flow problem:
\bc
\parbox{\hsize}{
\noindent
{\tt
{\bf algorithm} successive shortest paths (convex costs)\\
{\bf begin}\\
\spacea ${\bf x}:={\rm min}\,\{H({\bf x})\,|\,{\bf x}\in Z^A\}$ and $\pi:=0$\\
\spacea $e(i):=b(i)+\sum_{\{j|(ji)\in A\}} x_{ji}
-\sum_{\{j|(ji)\in A\}} x_{ij}$ $\forall i\in N$\\
\spacea {\bf while} there is a node $\s$ with $e(\s)>0$ {\bf do}\\
\spacea {\bf begin}\\
\spaceb compute the reduced costs $c^\pi({\bf x})$\\
\spaceb determine shortest path distance $d(\cdot)$ from $\s$ to\\
\spacec all other nodes in $G({\bf x})$ w.r.\ to the reduced costs $c_{ij}^\pi$\\
\spaceb choose a node $\t$ with $e(\t)<0$\\
\spaceb augment ${\bf x}$ along shortest path from $\s$ to $\t$ by {\it one} unit\\
\spaceb $\pi=\pi-d$\\
\spacea {\bf end}\\
{\bf end}
}}
\ec
Note that we start with an optimal solution for $H({\bf x})$, i.e.\ we
choose for each arc $(i,j)$ the value of $x_{ij}$ in such a way that
it is minimal. By this we guarantee that $c_{ij}(x_{ij})\ge0$ and thus
that the reduced costs $c_{ij}^\pi$ with $\pi=0$ fulfill the
optimality condition $c_{ij}^\pi\ge0$. The complexity of this
algorithm is strictly polynomial in the physical example we discuss in
section \ref{sos_section} \cite{blasum}.  \vskip2cm

\noindent
{\Large\bf Final remark}
\vskip0.5cm

For everybody who encounters one of the network flow problems
mentioned above in his study of a physical (or any other) problem an
important advice: Before reinventing the wheel, which means before
wasting the time in hacking a subroutine that solves a particular
network flow problem, one should consult the extremely valuable LEDA
(library of efficient data types and algorithms) library, where many
source codes of highly efficient combinatorial optimization algorithms
can be found. All information, the manual \cite{leda} and the source
codes can be found on the internet (this is public domain):
\bc
http://www.mpi--sb.mpg.de/LEDA/leda.html
\ec
Have fun!

\vfill 
\eject


\begin{thebibliography}

\bibitem{[1]}{toulouse}{[1]}
        G. Toulouse, Commun. Phys. {\bf 2}, 115 (1977).

\bibitem{[2]}{unfrust}{[2]}
        J. Villain, J. Phys. C {\bf 10}, 1717 (1977);
        G. Forgacs, Phys. Rev. B {\bf 22}, 4473 (1980);
        A. B. Harris, C. Kallin and A. J. Berlinsky,
        Phys. Rev. B {\bf 45}, 2899 (1992);
        D. A. Huse and A. D. Rutenberg,
        Phys. Rev. B {\bf 45}, 7536 (1992);
        J. T. Chalker, P. C. W. Holdsworth and E. F. Shender,
        Phys. Rev. Lett. {\bf 68}, 855 (1992);
        J. D. Shore, M . Holzer and J. P. Sethna,
        Phys. Rev. B {\bf 46}, 11376 (1992);
        P. Chandra, P. Coleman and I. Ritchey,
        J. Physique I {\bf 3}, 591 (1993);
        J. N. Reimers and A. J. Berlinsky,
        Phys. Rev. B {\bf 48}, 9539 (1993).
       

\bibitem{[3]}{tsp}{[3]}
        {\it The traveling salesman problem}, ed.:
        E. L. Lawler, J. K. Lenstra, A. H. G. Rinnooy Kan,
        D. B. Shmoys, Wiley-Interscience series in discrete mathematics,
        (John Wiley \& Sons, Chichester, 1985).

\bibitem{[4]}{wilson_bondy}{[4]}
        R. J. Wilson, {\it Introduction to graph theory},
        (Oliver and Boyd, Edinburgh, 1972);
        J. A. Bondy and U. S. R. Murty,
        {\it Graph theory with applications}, 
        (Mac Millan, London, 1976).

\bibitem{[5]}{lawler}{[5]}
        E. L. Lawler, 
        {\it Combinatorial optimization: Networks and matroids},
        (Holt, Rinehart and Winston, New York, 1976).

\bibitem{[6]}{papa}{[6]}
        C. H. Papadimitriou and K. Steiglitz,
        {\it Combinatorial optimization: Algorithms and complexity},
        (Prentice-Hall, Englewood Cliffs NJ, 1982).

\bibitem{[7]}{chvatal}{[7]}
        V. Chv\'atal, {\it Linear programming},
        (Freeman, San Francisco, 1983).

\bibitem{[8]}{derigs}{[8]}
        U. Derigs, {\it Programming in networks and graphs},
        Lecture Notes in Economics and Mathematical Systems {\bf 300},
        (Springer-Verlag, Berlin-Heidelberg, 1988).

\bibitem{[9]}{groetschel2}{[9]}
        M. Gr\"otschel, L. Lov\'asz and A. Schrijver,
        {\it Geometric algorithms and combinatorial optimization},
        (Springer-Verlag, Berlin-Heidelberg, 1988).

\bibitem{[10]}{ahuja}{[10]}
        R. K. Ahuja, T. L. Magnati and J. B. Orlin,
        {\it Network Flows}, (Prentice Hall, London, 1993).
        
\bibitem{[11]}{middleton}{[11]}
        A. A. Middleton, Phys. Rev. E {\bf 52}, R3337 (1995).

\bibitem{[12]}{rieger_review}{[12]}
        H.\ Rieger, {\it Monte Carlo simulations of Ising
        spin glasses and random field systems} in {\it Annual Reviews
        of Computational Physics II}, p.\ 295--341, (World Scientific,
        Singapore, 1995).

\bibitem{[13]}{barahona_rfim}{[13]}
        F.\ Barahona, J.\ Phys.\ A {\bf 18}, L673 (1985).

\bibitem{[14]}{ogielski}{[14]}
        A. T. Ogielski, Phys. Rev. Lett. {\bf 57}, 1251 (1986).

\bibitem{[15]}{fishman} {[15]}
        S. Fishman and A. Aharony, J. Phys. C {\bf 12}, L729 (1979).

\bibitem{[16]}{hartmann}{[16]}
        A. K. Hartmann and K. D. Usadel,
        Physica A {\bf 214}, 141 (1995).

\bibitem{[17]}{esser}{[17]}
        J. Esser, private communication.

\bibitem{[18]}{shklovsky}{[18]}
        A. L. Efros and B. I. Shklovskii,
        J.\ Phys.\ A {\bf 8}, L49 (1975).       

\bibitem{[19]}{tenelsen}{[19]}
        K.\ Tenelsen, M.\ Schreiber, 
        Phys.\ Rev.\ B {\bf 49}, 12622 (1994); {\bf 52}, 13287 (1995).

\bibitem{[20]}{barahona}{[20]}
        F.\ Barahona, J.\ Phys.\ A {\bf 15}, 3241 (1982);
        F.\ Barahona, R.\ Maynard, R.\ Rammal and J.\ P.\ Uhry,
        J.\ Phys.\ A {\bf 15}, 673 (1982).
        
\bibitem{[21]}{kawashima}{[21]}
        N. Kawashima and H. Rieger,
        submitted to Europhys. Lett., cond-mat/9612116.

\bibitem{[22]}{groetschel}{[22]}
        M.~Gr\"otschel, M.\ J\"unger and G.\ Reinelt,
        in: {\em Heidelberg Colloqium on Glassy dynamics and
        Optimization} , ed. L.\ van Hemmen and I.\ Morgenstern 
        (Springer-Verlag, Heidelberg 1985).

\bibitem{[23]}{juenger}{[23]}
        C. De Simone, M. Diehl, M. J\"unger, P. Mutzel, G. Reinelt 
        and G. Rinaldi, J.~Stat.~Phys. {\bf 80}, 487 (1995).

\bibitem{[24]}{kobe}{[24]}
        T. Klotz and S. Kobe,
        J. Phys. A {\bf 27}, L95 (1994).

\bibitem{[25]}{diehl}{[25]} 
        M.\ Diehl, {\it Determination of exact
        ground states of Ising spin glasses with a branch-and-cut
        algorithm}, (Diploma Thesis, K\"oln, 1995) unpublished,
        postscript file available at 
        http://www.informatik.uni-koeln.de/ls\_juenger/staff/diehl.html.

\bibitem{[26]}{thienel}{[26]}
        S. Thienel, {\it ABACUS --- A Branch--And-CUt System},
        (Ph.D. thesis, K\"oln, 1995) unpublished, postscript file
        available at 
        http://www.informatik.uni-koeln.de/ls\_juenger/publications/thienel/diss.html.

\bibitem{[27]}{bray} {[27]}
        A.~J.~Bray and M.~A.~Moore
        in: {\em Heidelberg Colloqium on Glassy dynamics and
        Optimization}, ed. L.\ van Hemmen and I.\ Morgenstern
        (Springer-Verlag, Berlin-Heidelberg 1985).

\bibitem{[28]}{rieger_sg}{[28]}
        H. Rieger, L. Santen, U. Blasum, M. Diehl, M. J\"unger and G. Rinaldi,
        J.\ Phys.\ A {\bf 29}, 3939 (1996).

\bibitem{[29]}{antiphase}{[29]}
         R.~Maynard and R.~Rammal, 
         J. Phys. Lett.(France) {\bf 43} L347(1982);
         Y. Ozeki,  J. Phys. Soc. Jpn. {\bf 59} 3531 (1990).

\bibitem{[30]}{siterandom}{[30]}
        T.~Shirakura and F.~Matsubara,
        J. Phys. Soc. Jpn.{\bf 64} 2338 (1995);
        Y.~Ozeki and Y.~Nonomura,
        J.~Phys.~Soc.~Jpn. {\bf 64} 3128 (1995).

\bibitem{[31]}{blasum}{[31]}
        U.\ Blasum, W.~Hochst{\"a}ttler, C.~Moll, H.\ Rieger,  
        J.\ Phys.\ A {\bf 29}, L459 (1996).

\bibitem{[32]}{rieger_sos}{[32]}
        H. Rieger and U. Blasum,
        Phys. Rev. Lett., in press.

\bibitem{[33]}{tsai}{[33]}
        Y.-C.~Tsai and Y.~Shapir, Phys.~Rev.~Lett. {\bf 69} (1992) 1773;
        Phys.~Rev.\ E {\bf 40} (194) 3546, 4445.

\bibitem{[34]}{puresos}{[34]}
        See e.\ g.\ S.~T.~Chui and J.~D.~Weeks,
        Phys.~Rev.\ B {\bf 14}, 4978 (1976).

\bibitem{[35]}{zeng}{[35]}
        C. Zeng, A. A. Middleton and Y. Shapir,
        Phys. Rev. Lett. {\bf 77}, 3204 (1996).

\bibitem{[36]}{kleinert}{[36]}
        H.\ Kleinert, {\em Gauge Fields in Condensed Matters}, 
        (World Scientific, Singapore, 1989).

\bibitem{[37]}{rieger_vortex}{[37]}
        H. Rieger and J. Kisker, in preparation.

\bibitem{[38]}{wengel}{[38]}
        C.~Wengel and A.~P.~Young,  
        Phys.~Rev.~B {\bf 54}, R6869 (1996).

\bibitem{[39]}{bokil}{[39]}
        H. S. Bokil and A. P. Young,
        Phys. Rev. Lett. {\bf 74}, 3021 (1995).

\bibitem{[40]}{traffic}{[40]}
        {\it Traffic and Granular Flow}, ed. D. E. Wolf,
        M. Schreckenberg and A. Bachem (World Scientific, Singapore 1996).

\bibitem{[41]}{leda}{[41]}
        S. N\"aher and C. Uhrig, 
        {\it The LEDA User Manual, Version R3.4}
        (Martin--Luther Universit\"at Halle--Wittenberg, Germany,
        1996), postscript file available at 
        http://www.mpi-sb.mpg.de/LEDA/leda.html.


\end{thebibliography}
\end{document}